\documentclass[a4paper,11pt]{article}
\usepackage{jheppub} 
\usepackage{color}
\usepackage{comment}
\usepackage{amsmath,amssymb,mathtools}
\usepackage{cancel}
\usepackage{ascmac}
\usepackage{enumerate}
\usepackage[all]{xy}
\usepackage{listings}
\usepackage{caption}
\usepackage{here}
\usepackage{braket}
\usepackage{tcolorbox}
\usepackage{tikz-feynhand}

\newcommand{\KFmod}[1]{{#1}}
\newcommand{\SYmod}[1]{{#1}}

\title{\boldmath Magnetic Catalysis and Fermion Mass Generation in de Sitter Spacetime}

\author[a]{Kohei Fujikura,}
\author[b,c]{Toshifumi Noumi,}
\author[b]{and Shintetsu Yamazaki}

\affiliation[a]{Yukawa Institute for Theoretical Physics, Kyoto University, Kyoto 606-8502, Japan}

\affiliation[b]{Graduate School of Arts and Sciences, University of Tokyo, Komaba, Meguro-ku, Tokyo 153-8902, Japan}

\affiliation[c]{RIKEN Center for Interdisciplinary Theoretical and Mathematical Sciences (iTHEMS), RIKEN, Wako 351-0198, Japan}

\emailAdd{kohei.fujikura@yukawa.kyoto-u.ac.jp,
tnoumi@g.ecc.u-tokyo.ac.jp, shintetsuperro630@g.ecc.u-tokyo.ac.jp}

\abstract{
We consider the dynamics of a charged fermion in de Sitter space in the presence of a uniform background magnetic field, and discuss magnetic catalysis of chiral symmetry breaking using the Nambu--Jona-Lasinio (NJL) model.
We evaluate the mode functions of the charged fermion field in this background by imposing the Bunch--Davies vacuum condition.
The gap equation is solved in the mean-field approximation.
We derive analytic expressions for the gap in several limiting regimes, such as the large-magnetic-field and large-curvature limits.
We find that the curvature effect restores chiral symmetry, whereas the magnetic field enhances chiral symmetry breaking through the conventional mechanism of magnetic catalysis.
The phase structure associated with chiral symmetry breaking is revealed by numerical calculations.
}

\def \b{\boldsymbol}

\begin{document}

\begin{flushright}
\small RIKEN-iTHEMS-Report-26, YITP-26-108
\end{flushright}
\flushbottom
\maketitle

\section{Introduction}

The origin of primordial magnetic fields is an important open problem in cosmology~\cite{Durrer:2013pga,Kandus:2010nw}.
Among various proposed mechanisms, inflationary electromagnetic field production is particularly attractive because it can generate magnetic fields with large correlation lengths, which may survive until the present epoch~\cite{Turner:1987bw,Ratra:1991bn,Demozzi:2009fu,Anber:2009ua,Barnaby:2011qe,Adshead:2016iae,Talebian:2021dfq}.
This motivates a basic field-theoretic study of charged light fermions in inflationary magnetic backgrounds, which may provide a useful basis for discussing model-dependent cosmological applications.\footnote{For possible cosmological implications, see, e.g.,~\cite{Giovannini:1997eg,Giovannini:1997gp,Kamada:2016eeb,Domcke:2019mnd,Kamada:2020bmb,Yanagihara:2023qvx,Di:2024gsl,Boyer:2025jno,Hamada:2025cwu,Hamada:2025ooy,Fukuda:2025nmc}; for gravitational waves sourced by electromagnetic fields, see~\cite{Caprini:2014mja,Sharma:2019jtb,Okano:2020uyr,Brandenburg:2021bfx,Maiti:2024nhv,Atkins:2025pvg}.}

A charged light fermion in an inflationary universe under an external electromagnetic field exhibits several nonperturbative phenomena, such as chiral fermion production~\cite{Domcke:2018eki,Domcke:2019qmm}, induced electric currents including the chiral magnetic effect~\cite{Gorbar:2021ajq,Gorbar:2023zla}, and the Schwinger effect in an inflationary universe~\cite{Kobayashi:2014zza,Sobol:2018djj,Sobol:2019xls,Gorbar:2021zlr,Cado:2022pxk}.
However, the dynamics of charged fermions in the presence of a background magnetic field in de Sitter space is not fully understood yet.
In particular, these phenomena are sensitive to an infrared property of charged fermions such as the fermion mass.
Hence, it is crucial to understand the effect of the background magnetic field on charged fermions from various perspectives.

In flat space, it is known that a background magnetic field catalyzes chiral symmetry breaking~\cite{Gusynin:1994re,Gusynin:1994va}.
An essential observation is that the motion of charged particles is restricted to the direction parallel to the uniform magnetic field, and the dynamics of the lowest Landau level (LLL) is effectively described by a $(1+1)$-dimensional theory in the large-magnetic-field limit.
Consequently, chiral symmetry breaking is catalyzed even by an infinitesimally weak attractive interaction at zero temperature.
This phenomenon, known as magnetic catalysis, has been extensively investigated in various contexts, including QCD in large magnetic fields at both zero and finite temperature~\cite{Shovkovy:2012zn,Kharzeev:2012ph,Miransky:2015ava,Andersen:2014xxa,Mizher:2010zb,Fukushima:2010fe,Bali:2011qj,Bali:2012zg}.
An enhancement of the chiral condensate in a uniform magnetic field with an attractive interaction is interesting because a charged fermion in an inflationary universe can acquire a mass, which may affect the infrared properties of the phenomena mentioned above.

However, the situation is less obvious in an inflationary universe, which is well approximated by de Sitter space.
de Sitter space is a maximally symmetric solution of the Einstein equations with a positive cosmological constant and is characterized by the Hubble scale $H$.
The seminal work of Gibbons and Hawking showed that quantum field theory in de Sitter space admits a thermal interpretation with temperature $T_{\mathrm{dS}}=H/(2\pi)$~\cite{Gibbons:1977mu}.
Therefore, the curvature effect restores chiral symmetry, opposing magnetic catalysis, if this naive interpretation persists.
The central question of this work is whether magnetic catalysis can operate in de Sitter space.
For generic spacetimes, this question has also been addressed in refs.~\cite{Geyer:1996np,Flachi:2015sva} within the small-curvature approximation, with results suggesting curvature-induced chiral symmetry restoration.

In this paper, we revisit this question in de Sitter space.
We consider the Nambu--Jona-Lasinio (NJL) model~\cite{Nambu:1961tp,Nambu:1961fr} in de Sitter space in the presence of a uniform background magnetic field without making the small-curvature approximation.
We quantize the charged fermion field in de Sitter space by imposing the Bunch--Davies vacuum condition~\cite{Bunch:1978yq}, and solve the gap equation to reveal the phase structure. See also refs.~\cite{Hashida:1999wb,Tong:2023krn,Fujikura:2025xgl} for recent developments on the NJL model in an inflationary universe without an external magnetic field.

We find analytic expressions for the effective potential of the chiral condensate in the limits of a large magnetic field and a large Hubble parameter within the mean-field approximation.
It turns out that the Hubble expansion restores chiral symmetry, whereas the magnetic field enhances chiral symmetry breaking through the conventional mechanism of magnetic catalysis.
For generic parameter values, we solve the gap equation numerically and reveal the phase structure.
Our numerical results show that there is a continuous phase transition associated with restoration (or breaking) of chiral symmetry as the strength of the magnetic field or the Hubble parameter is varied.
We also discuss the limitations of the mean-field NJL analysis and explain why more sophisticated approaches beyond the NJL model are required before drawing phenomenological implications for cosmology.

This paper is organized as follows. In Sec.~\ref{sec:charged_fermion}, we derive mode functions of
charged fermions in de Sitter space in a uniform magnetic field,
evaluate the chiral condensate, and introduce the NJL model.
In Sec.~\ref{sec:phase_diagram}, we solve the gap equation and discuss the resulting phase diagram.
Sec.~\ref{sec:conclusion} is devoted to
conclusions and discussion.

\paragraph{Notation and conventions.}

Throughout this paper, we adopt the following notation and conventions.
We use natural units $\hbar=c=k_B=1$.
We work in the spatially flat FLRW coordinates
$x^\mu=(\eta,\boldsymbol{x})$, where $\eta$ is the conformal time and
$\boldsymbol{x}=(x,y,z)$ denotes the comoving spatial coordinates.
The de Sitter line element is written as
\begin{align}
    \mathrm{d}s^2
    =
    a^2(\eta)
    \left(
        \mathrm{d}\eta^2
        -
        \mathrm{d}\boldsymbol{x}^2
    \right),
    \label{eq:metric_de_sitter}
\end{align}
where
\begin{equation}
    a(\eta)
    =
    -\frac{1}{H\eta}
\end{equation}
is the scale factor and $H$ is the Hubble parameter.

Greek indices $\mu,\nu,\ldots$ label curved-spacetime components and
take values $0,1,2,3$, whereas Latin indices $a,b,\ldots$ label
components in the local Lorentz frame and likewise take values
$0,1,2,3$. Spatial indices $i,j,k,\ldots$ take values $1,2,3$, and the
spatial coordinates are written as
\begin{equation}
    x^i=(x,y,z).
\end{equation}

We denote by $\gamma^a$ the standard gamma matrices defined in the
local Lorentz frame. They satisfy the Clifford algebra
\begin{equation}
    \{\gamma^a,\gamma^b\}
    =
    2\eta^{ab}I,
    \qquad
    \eta^{ab}
    =
    \mathrm{diag}(+1,-1,-1,-1),
\end{equation}
where $I$ denotes the $4\times4$ identity matrix in spinor space.
In the Dirac representation, the gamma matrices are explicitly given by
\begin{align}
    \gamma^0
    &=
    \begin{pmatrix}
        I & 0 \\
        0 & -I
    \end{pmatrix},
    &
    \gamma^i
    &=
    \begin{pmatrix}
        0 & \sigma^i \\
        -\sigma^i & 0
    \end{pmatrix},
\end{align}
where $I$ in these block matrices denotes the $2\times2$ identity
matrix, and $\sigma^i$ is the $i$-th Pauli matrix. The Pauli matrices
are explicitly given by
\begin{align}
    \sigma^1
    &=
    \begin{pmatrix}
        0 & 1 \\
        1 & 0
    \end{pmatrix},
    &
    \sigma^2
    &=
    \begin{pmatrix}
        0 & -i \\
        i & 0
    \end{pmatrix},
    &
    \sigma^3
    &=
    \begin{pmatrix}
        1 & 0 \\
        0 & -1
    \end{pmatrix}.
\end{align}

The curved-spacetime gamma matrices $\hat{\gamma}^{\mu}$ are defined
in terms of the tetrad (vielbein) $e^\mu{}_a$ as
\begin{equation}
    \hat{\gamma}^{\mu}
    =
    e^\mu{}_a\gamma^a,
    \qquad
    e^\mu{}_a
    =
    \frac{\delta^\mu_a}{a(\eta)}.
\end{equation}

\section{Charged fermion in de Sitter space and Nambu--Jona-Lasinio model}\label{sec:charged_fermion}

In this section, we study the chiral condensate in the NJL model for a
single Dirac fermion in de Sitter space in the presence of a
homogeneous external magnetic field satisfying the source-free Maxwell
equations. We take the comoving magnetic field to be constant.

We first quantize a free massive charged fermion in the Bunch--Davies vacuum, derive its mode functions, and evaluate the regularized fermion bilinear $\langle\bar\psi\psi\rangle$.
We then study some extreme limits including a large magnetic field and a large Hubble parameter to isolate the effects of curvature and the magnetic field.
Finally, we introduce the NJL model.

\subsection{A charged fermion in de Sitter space with a uniform magnetic field}
\label{sec:charged_fermion_uniform_B}

In this subsection, we consider a free massive Dirac fermion minimally coupled to gravity
and to an external magnetic field in de Sitter space. In the
spatially flat FLRW coordinates $(\eta,\boldsymbol{x})$ introduced
above, its action is given by
\begin{align}
S = \int \mathrm{d}^4x\sqrt{-g}\left[i\bar{\psi}\hat{\gamma}^\mu D_\mu\psi - m\bar{\psi}\psi\right].
\end{align}
Here we set $D_\mu=\partial_\mu-ieA_\mu+\Gamma_\mu$, where $\Gamma_\mu$ and $A_\mu$ are the spin connection and the background gauge field, respectively.

    In the following, we derive an explicit form of mode functions. We next evaluate the fermion bilinear $\langle\bar{\psi}\psi\rangle$ and summarize the regularization scheme.
    These results will be used to discuss the generation of a fermion mass in the NJL model of Sec.~\ref{sec:njl_model}.

\subsubsection{Mode functions}

To quantize the charged fermion, let us begin with the Dirac equation,
\begin{align}
&(i\hat{\gamma}^\mu D_\mu-m)\psi(x)=0.\label{eq:dirac_equation}
\end{align}
An explicit form of the spin connection is given by
\begin{align}
\Gamma_0=0,\qquad \Gamma_i=-\dfrac{1}{2}Ha\gamma^i\gamma^0\quad (i=1,2,3).
\end{align}
The Dirac equation can be simplified by introducing the rescaled spinor field $\xi = a^{3/2}(\eta)\psi$ as
\begin{equation}
    (i\gamma^\mu(\partial_\mu - ieA_\mu) - ma(\eta))\xi = 0.\label{eq:simplified_dirac_equation}
\end{equation}
We adopt the gauge $A_\mu=(0,-By,0,0)$, which corresponds to a uniform background magnetic field $B>0$ directed along the $z$-axis.
\KFmod{Here $B$ is the constant comoving magnetic field, corresponding to the physical field $B_{\mathrm{phys}}(\eta)=B/a^2(\eta)$. The mode functions below are derived for this background and a constant fermion mass $m$.}
In the Poincar\'e patch \eqref{eq:metric_de_sitter}, the structure of the above equation is the same as that in flat space, except for the temporal part. \SYmod{We take the gauge coupling $e$ to be positive here.}

We decompose the Dirac spinor as $\xi = (\phi, \chi)^T$, where $\phi$ and $\chi$ are upper and lower two-component spinors. Eq.~\eqref{eq:simplified_dirac_equation} is expressed in block matrix form as
\begin{align}
    \begin{pmatrix}
        i\partial_0 - ma(\eta) & i\b{\sigma}\cdot\b{\partial} - eB\sigma^1 y \\ 
        -i\b{\sigma}\cdot\b{\partial} + eB\sigma^1 y & -i\partial_0 - ma(\eta)
    \end{pmatrix}
    \begin{pmatrix}
        \phi \\ \chi 
    \end{pmatrix}
    &= 0. \label{eq:dirac_equation_block}
\end{align}
It follows from the above expression that $\phi$ satisfies the following second-order differential equation:
\begin{equation}
    \left[\partial_0^2 + ima^2H\left(1-i\dfrac{m}{H}\right)- (\partial_x+ieBy)^2-\partial_y^2-\partial_z^2 - eB\sigma^3\right]\phi=0.\label{eq:eom_phi}
\end{equation}

Motivated by the analogy with the mode functions in flat space, we consider the following mode expansion for the fermion field:
\begin{align}
    \phi(\eta,\b{x})
    &=
    \sum_{N,s} \int\frac{\mathrm{d}^2\bar{\b{k}}}{(2\pi)^2}
    \Bigg[
    a_{\bar{\b{k}},s,N}h_N(\bar y)e^{ik_x x+ik_z z}f^{(+)}_{N,s,\bar{\b{k}}}(\eta)
    \nonumber\\*
    &\hspace{35mm}
    + b^{\dagger}_{\bar{\b{k}},s,N}h_N(\bar y_-)e^{-ik_x x-ik_z z}{f}^{(-)}_{N,s,\bar{\b{k}}}(\eta)
    \Bigg].\label{eq:mode_expansion}
\end{align}
Here, $s=\pm 1$ denotes the spin and $N=0,1,2,\ldots$ is a nonnegative integer labeling the Landau levels, and we define $\bar{\b{k}}=(k_x,k_z)$ and $\mathrm{d}^2\bar{\b{k}}=\mathrm{d}k_x\,\mathrm{d}k_z$.
In particular, $\sigma^3f^{(\pm)}_s =sf^{(\pm)}_s$ is satisfied.
$h_N$ is the square-normalizable mode function for the $y$-direction, and is given by
\begin{align}
    &h_N(u)=\left(\dfrac{\sqrt{eB}}{2^N\sqrt\pi N!}\right)^{1/2}e^{-u^2/2}\tilde{h}_N(u),\\
    &\bar{y}=\sqrt{eB}y+\dfrac{k_x}{\sqrt{eB}},\quad
    \bar{y}_-=\sqrt{eB}y-\dfrac{k_x}{\sqrt{eB}} ,
\end{align}
where $\tilde{h}_N$ is the Hermite polynomial.
The mode functions $f^{(+)}_{N,s}(\eta)$ and $f^{(-)}_{N,s}(\eta)$ are chosen so that they approach positive- and negative-frequency solutions in the remote past $\eta \to -\infty$.
The operators $a_{\bar{\b{k}},s,N}$ and $b_{\bar{\b{k}},s,N}$ annihilate particles and antiparticles, respectively.
The Bunch--Davies vacuum $|{\mathrm{BD}}\rangle$ is defined by the relation,
\begin{align}
    a_{\bar{\b{k}},s,N}|{\mathrm{BD}}\rangle = b_{\bar{\b{k}},s,N}|{\mathrm{BD}}\rangle = 0, \quad \forall \,\bar{\b{k}}, N, s.
\end{align}
The effect of spacetime curvature is encoded in $f_s(\eta)$.

Putting the mode expansion \eqref{eq:mode_expansion} into Eq.~\eqref{eq:eom_phi}, one obtains the differential equation for $f^{(\pm)}_s(\eta)$ as
\begin{align}
    &\left(\partial_\eta^2 + \frac{im}{H\eta^2}\left(1-i\dfrac{m}{H}\right) + \Omega^2_{N,s}\right)f^{(\pm)}_{N,s,\bar{\b{k}}}(\eta) = 0,~
    \Omega_{N,s}^2=k_z^2 + (2N+1-s)eB.
\end{align}
A solution of the above equation reads
\begin{align}
f^{(+)}_{N,s,\bar{\b{k}}}(\eta)=C_{+}\mathcal{H}^{(1)}_\nu(-\Omega_{N,s}\eta)f_s^{(+)},~f^{(-)}_{N,s,\bar{\b{k}}}(\eta)=C_{-}\mathcal{H}^{(2)}_\nu(-\Omega_{N,s}\eta)f_s^{(-)},
\end{align}
where we defined
\begin{align}
    \mathcal{H}_\nu^{(1,2)}(z)=\sqrt{z}H^{(1,2)}_\nu(z),
    \quad
    \nu = \dfrac{1}{2}-i\dfrac{m}{H}
    .
\end{align}
Also, $C_{\pm}$ are constants that can be fixed by the canonical anticommutation relation.
$f_{N,s,\bar{\b{k}}}(\eta)$ is labeled by the Landau level $\Omega_{N,s}$.
There are two degenerate modes, $\Omega_{N,s=-1}=\Omega_{N+1,s=+1}$, except for the zero mode corresponding to $N=0$ and $s=1$.
This structure is the same as that in flat space.
To account for this degeneracy, it is convenient to introduce the Landau-level index $n=N+(1-s)/2$.

$\chi(\eta,\b{x})$ can be determined from $\phi$ by the original Dirac equation \eqref{eq:dirac_equation_block}.
Consequently, one obtains the following mode expansion:
\begin{align}
    \psi(\eta,\b{x}) = \int\frac{\mathrm{d}^{2}\bar{\b{k}}}{(2\pi)^2} \sum_{n,s}\Bigg[a_{\bar{\b{k}},s,n}U_{n,s,\bar{\b{k}}}(\eta,\b{x}) + b^{\dagger}_{\bar{\b{k}},s,n}V_{n,s,\bar{\b{k}}}(\eta,\b{x})\Bigg].
\end{align}
\SYmod{The mode sum contains $s=+1$ for $n=0$ and $s=\pm1$ for $n\geq1$, as follows from $N=n-(1-s)/2\geq0$. }Here, the mode functions $U_{n,s,\bar{\b{k}}}$ and $V_{n,s,\bar{\b{k}}}$ with $N=n-(1-s)/2$ are given by
\begin{align}
U_{n,s=+}(\eta,\b{x})
&=
\frac{e^{ik_x x + ik_z z}}{C_n\,a^{3/2}(\eta)}
\begin{pmatrix}
\mathcal{H}_{\nu}^{(1)}(-\omega_n\eta)\,h_N(\bar y) \\
0 \\
-\,i\,\dfrac{k_z}{\omega_n}\,\mathcal{H}_{\nu-1}^{(1)}(-\omega_n\eta)\,h_N(\bar y) \\
-\,i\,\dfrac{\sqrt{2neB}}{\omega_n}\,\mathcal{H}_{\nu-1}^{(1)}(-\omega_n\eta)\,h_{N-1}(\bar y)
\end{pmatrix},
\label{eq:mode_function_1}\\[1em]
U_{n,s=-}(\eta,\b{x})
&=
\frac{e^{ik_x x + ik_z z}}{C_n\,a^{3/2}(\eta)}
\begin{pmatrix}
0 \\
\mathcal{H}_{\nu}^{(1)}(-\omega_n\eta)\,h_N(\bar y) \\
-\,i\,\dfrac{\sqrt{2neB}}{\omega_n}\,\mathcal{H}_{\nu-1}^{(1)}(-\omega_n\eta)\,h_{N+1}(\bar y) \\
+\,i\,\dfrac{k_z}{\omega_n}\,\mathcal{H}_{\nu-1}^{(1)}(-\omega_n\eta)\,h_N(\bar y)
\end{pmatrix},\label{eq:mode_function_2}\\
V_{n,s=+}(\eta,\b{x})
&=
\frac{e^{-ik_x x - ik_z z}}{C_n\,a^{3/2}(\eta)}
\begin{pmatrix}
-\,i\,\dfrac{k_z}{\omega_n}\,\mathcal{H}_{\nu^*-1}^{(2)}(-\omega_n\eta)\,h_N(\bar y_-) \\
+\,i\,\dfrac{\sqrt{2neB}}{\omega_n}\,\mathcal{H}_{\nu^*-1}^{(2)}(-\omega_n\eta)\,h_{N-1}(\bar y_-) \\
-\,\mathcal{H}_{\nu^*}^{(2)}(-\omega_n\eta)\,h_N(\bar y_-) \\
0
\end{pmatrix},
\label{eq:mode_function_3}\\[1em]
V_{n,s=-}(\eta,\b{x})
&=
\frac{e^{-ik_x x - ik_z z}}{C_n\,a^{3/2}(\eta)}
\begin{pmatrix}
+\,i\,\dfrac{\sqrt{2neB}}{\omega_n}\,\mathcal{H}_{\nu^*-1}^{(2)}(-\omega_n\eta)\,h_{N+1}(\bar y_-) \\
+\,i\,\dfrac{k_z}{\omega_n}\,\mathcal{H}_{\nu^*-1}^{(2)}(-\omega_n\eta)\,h_N(\bar y_-) \\
0 \\
-\,\mathcal{H}_{\nu^*}^{(2)}(-\omega_n\eta)\,h_N(\bar y_-)
\end{pmatrix}\label{eq:mode_function_4},
\end{align}
where we introduced
\begin{align}
\omega_n^2 = k_z^2 + 2neB.
\end{align}
The normalization constant $C_n$ is fixed by the canonical anticommutation relation as
\begin{equation}
    C_n=\sqrt{\dfrac{4}{\pi}}e^{-\pi m/(2H)}.
\end{equation}
For $n=0$, we define $h_{-1}=0$.

\subsubsection{Fermion bilinear}

Next we calculate the fermion bilinear using the obtained mode functions.
The composite operator $\bar{\psi}(\eta,\b{x})\psi(\eta,\b{x})$ is defined by point splitting as
\begin{align}
\bar{\psi}(\eta,\b{x})\psi(\eta,\b{x})
\equiv
\lim_{\b{x}_1 \to \b{x}_2}
\bar{\psi}(\eta,\b{x}_1)U(\b{x}_1,\b{x}_2)\psi(\eta,\b{x}_2),
\end{align}
where $U(\b{x}_1,\b{x}_2)$ is the straight Wilson line connecting $\b{x}_1$ and $\b{x}_2$.
We can evaluate the chiral condensate in the free theory as
\begin{align}
    &\langle {\mathrm{BD}}|\bar{\psi}(\eta,\b{x})\psi(\eta,\b{x}) |{\mathrm{BD}}\rangle = H^3\int^\infty_{-\infty} \frac{\mathrm{d}\tilde{k}_z}{2\pi} \dfrac{e^{\pi \tilde{m}} e\tilde{B}}{4}\sum_{n=0}^{\infty} \mathcal{I}_{n}(\tilde{k}_z,\tilde{B}),\\
    &\mathcal{I}_n(\tilde{k}_z,\tilde{B})=
\begin{dcases}
\dfrac{\tilde{\omega}_{0}}{2}(|H^{(2)}_{\nu^*-1}(\tilde{\omega}_0)|^2-|H^{(2)}_{\nu^*}(\tilde{\omega}_0)|^2) & (n=0),\\
\tilde{\omega}_n(|H^{(2)}_{\nu^*-1}(\tilde{\omega}_n)|^2-|H^{(2)}_{\nu^*}(\tilde{\omega}_n)|^2) & (n\geq1),
\end{dcases}\label{eq:chiral_condensate_one_loop}\\
&\tilde{B}=B\eta^2,~\tilde{k}_z=-k_z\eta,~\tilde{\omega}_n^2=\tilde{k}_z^2 + 2ne\tilde{B},~\tilde{m}=m/H.
\end{align}
The physical magnetic field and the physical momentum are given by $B_{\mathrm{phys}}=\tilde{B}H^2$ and $k_{{\mathrm{phys}}}=\tilde{k}_zH$, respectively.

We regularize the UV divergence of the above integral in the following two
steps: First, at each Landau level, we isolate from the
exact integrand the terms responsible for the UV divergences. The
remaining part is UV convergent and will be evaluated numerically in
Sec.~\ref{sec:numerical_phase_diagram}. Second, we evaluate the isolated UV part analytically using
a Schwinger proper-time cutoff $\Lambda$. The UV divergences arise
from the region $n\gg 1$ and $|\tilde{k}_z|\gg 1$, which corresponds to
the large-argument regime of the Hankel functions. Then, the terms to be isolated can be identified based on the following asymptotic
expansion:
\begin{align}
    z(|H^{(2)}_{\nu^*-1}(z)|^2-|H^{(2)}_{\nu^*}(z)|^2) &= e^{-\pi \tilde{m}}\left(-\frac{4}{\pi}\frac{\tilde{m}}{z}+\frac{2}{\pi}\left(1+\tilde{m}^2\right)\frac{\tilde{m}}{z^3}\right)+\mathcal{O}\left(\frac{1}{z^5}\right)\nonumber\\
    &=e^{-\pi \tilde{m}}\left(-\frac{4}{\pi}\frac{\tilde{m}}{\sqrt{z^2+\tilde{m}^2}}+\frac{2}{\pi}\frac{\tilde{m}}{(z^2+\tilde{m}^2)^{3/2}}\right)+\mathcal{O}\left(\frac{1}{z^5}\right).
\end{align}
In the second line, we changed the expansion parameter to avoid an unphysical IR divergence at $z=0$.
This choice of the IR regulator $\tilde{m}$ ensures that the UV divergent part is identical to that in flat space, as we will see below.
Using this relation, one can decompose the integrand into UV convergent and UV divergent parts as
\begin{align}
    \mathcal{I}_n &= \mathcal{I}_n^{\mathrm{con}}+\mathcal{I}_n^{\mathrm{div}},\\
    \mathcal{I}^{\mathrm{div}}_n&=
    \begin{dcases}
    \frac{e^{-\pi \tilde{m}}}{2}\left(-\frac{4}{\pi}\frac{\tilde{m}}{\tilde{E}_0}\right)\qquad\qquad\,\, (n=0),\\
    e^{-\pi \tilde{m}}\left(-\frac{4}{\pi}\frac{\tilde{m}}{\tilde{E}_n}+\frac{2}{\pi}\frac{\tilde{m}}{\tilde{E}^3_n}\right)\quad(n \geq 1),
    \end{dcases}\\
    \tilde{E}_n^2&=(\tilde{k}_z^2+2ne\tilde{B} +\tilde{m}^2),
\end{align}
where $\mathcal{I}_n^{\mathrm{con}}$ ($\mathcal{I}_n^{\mathrm{div}}$) is the UV convergent (UV divergent) part.
$\mathcal{I}_n(\tilde{k}_z,\tilde{B})$ is defined by \eqref{eq:chiral_condensate_one_loop}.
The chiral condensate is therefore divided into the UV convergent part $\langle\bar{\psi}\psi\rangle_{\mathrm{con}}$ and the UV divergent part $\langle \bar{\psi}\psi\rangle_{\Lambda}$ as
\begin{align}
    &\langle {\mathrm{BD}}|\bar{\psi}(\eta,\b{x})\psi(\eta,\b{x}) |{\mathrm{BD}}\rangle = \langle \bar{\psi}\psi\rangle_{\mathrm{con}} + \langle \bar{\psi}\psi\rangle_{\Lambda},\label{eq:renormalized_chiral_condensate}\\
    &\langle \bar{\psi}\psi\rangle_{\mathrm{con}} =  H^3\int^\infty_{-\infty} \frac{\mathrm{d}\tilde{k}_z}{2\pi} \dfrac{e^{\pi \tilde{m}} e\tilde{B}}{4}\sum_{n=0}^{\infty} \mathcal{I}^{\mathrm{con}}_{n}(\tilde{k}_z,\tilde{B}),\\
    &\langle \bar{\psi}\psi\rangle_{\Lambda} = H^3\int^\infty_{-\infty} \frac{\mathrm{d}\tilde{k}_z}{2\pi} \dfrac{e^{\pi \tilde{m}} e\tilde{B}}{4}\sum_{n=0}^{\infty} \mathcal{I}^{\mathrm{div}}_{n}(\tilde{k}_z,\tilde{B}).
\end{align}
Here, the subleading term in the $n=0$ contribution is UV convergent, and hence we do not include it in $\mathcal{I}_0^{\mathrm{div}}$.
There is no analytic expression for the UV convergent part for general values of the parameters, and we evaluate it numerically.
However, the convergence is guaranteed by construction, which makes the numerical evaluation stable.

The UV divergent part is regularized by the Schwinger proper-time method~\cite{Schwinger:1951nm}.
We use the following regularization,
\begin{align}
    \dfrac{1}{A^p}\to \dfrac{1}{\Gamma(p)}\int^\infty_{1/\Lambda^2}\mathrm{d}\tau\,\tau^{p-1}e^{-A\tau}, \label{eq:proper_time_regularization}
\end{align}
where $\Lambda$ is regarded as the physical UV cutoff scale when $A$ has mass dimension two.
In particular, $p=1/2$ and $p=3/2$ are relevant for the present case.
This regularization yields
\begin{align}
    &\int^\infty_{-\infty}\dfrac{\mathrm{d}\tilde{k}_z}{2\pi}\left(\dfrac{1}{2\tilde{E}_0^{2p}}+\sum_{n=1}^\infty\dfrac{1}{\tilde{E}_n^{2p}}\right)\to\dfrac{1}{4\sqrt \pi \,\Gamma(p)}\int^\infty_{1/\tilde{\Lambda}^2}\mathrm{d}\tau\,\tau^{p-\frac{3}{2}}e^{-\tilde{m}^2\tau}\coth(e\tilde{B}\tau),
    \\&\int^\infty_{-\infty}\dfrac{\mathrm{d}\tilde{k}_z}{2\pi}\sum_{n=1}^\infty\dfrac{1}{\tilde{E}_n^{2p}}\to\dfrac{1}{4\sqrt \pi \,\Gamma(p)}\int^\infty_{1/\tilde{\Lambda}^2}\mathrm{d}\tau\,\tau^{p-\frac{3}{2}}e^{-\tilde{m}^2\tau}\left(\coth(e\tilde{B}\tau)-1\right).
\end{align}
In this manipulation, we use \eqref{eq:proper_time_regularization}, and perform the series summation and integration over $\tilde{k}_z$.
$\tilde{\Lambda}=\Lambda/H$ is the cutoff scale normalized by the Hubble parameter.
Then, $\langle \bar{\psi}\psi \rangle_{\Lambda}$ is regularized as
\begin{align}
    &\langle \bar{\psi}\psi \rangle_\Lambda = \langle \bar{\psi}\psi\rangle_{\mathrm{flat}}+H^2m\dfrac{eB_{\mathrm{phys}}}{4\pi^2}\int^{\infty}_{1/\Lambda^2}\mathrm{d}\tau\,e^{-m^2\tau} \left(\coth(eB_{\mathrm{phys}}\tau)-1\right),\\*
    &\langle \bar{\psi}\psi\rangle_{\mathrm{flat}}=-m\frac{eB_{\mathrm{phys}}}{4\pi^2}\int^\infty_{1/\Lambda^2}\mathrm{d}\tau\,\frac{e^{-m^2\tau}}{\tau}\coth(eB_{\mathrm{phys}}\tau).
\end{align}
The leading contribution $\langle\bar{\psi}\psi\rangle_{\mathrm{flat}}$ is identical to the chiral condensate in flat space, which does not depend on $H$ in physical units.
The subleading contribution is logarithmically divergent and is proportional to $H^2$ when the fermion mass, the UV cutoff, and the magnetic field are kept fixed in physical units.
Thus, our regularization scheme matches the usual flat-space proper-time regularization.
We treat $\Lambda$ as a physical cutoff scale, which is regarded as a model parameter.

\subsection{Some extreme limits of the chiral condensate}\label{sec:chiral_condensate_limits}

Before introducing the NJL model, we study the behavior of the fermion bilinear in some extreme limits to isolate the effects of the magnetic field and curvature. Moreover, simple closed-form expressions are available in these limits and are useful for understanding its physical behavior.

We first consider the case in which the magnetic field is small compared with the Hubble scale, $e\tilde{B}\ll 1$.
In this limit, $\langle \bar{\psi}\psi\rangle_\Lambda$ can be computed straightforwardly using the expansion $\coth x=1/x+\mathcal{O}(x)$, and one finds
\begin{align}
    &\langle \bar{\psi}\psi\rangle_\Lambda = -\dfrac{m}{4\pi^2}\left(\Lambda^2e^{-m^2/\Lambda^2}-m^2\Gamma(0,m^2/\Lambda^2)-H^2\Gamma(0,m^2/\Lambda^2)\right)+\mathcal{O}(eB_{\mathrm{phys}}), \label{eq:vanishing_magnetic_field_limit_divergent}
\end{align}
where $\Gamma(0,x)$ is the incomplete gamma function defined by
\begin{align}
    \Gamma(0,x)\coloneqq \int_x^\infty \mathrm{d}t\,\frac{e^{-t}}{t}.
\end{align}
The convergent part also takes a simple analytic form in this limit:
\begin{align}
\langle\bar{\psi}\psi \rangle_{\mathrm{con}}=
\frac{H^3}{4\pi^2}\tilde{m}\Bigl[
1
+2(1+\tilde{m}^2)\log\tilde{m}
-(1+\tilde{m}^2)\bigl(\psi^{(0)}(-1-i\tilde{m})+\psi^{(0)}(-1+i\tilde{m})\bigr)
\Bigr]. \label{eq:vanishing_magnetic_field_limit_convergent}
\end{align}
Here, $\psi^{(0)}(z)$ is the digamma function defined by $\psi^{(0)}(z)\coloneqq\Gamma'(z)/\Gamma(z)$, where $\Gamma(z)$ is the standard gamma function.
We derive this expression in Appendix~\ref{app:chiral_condensate}.

For $H\to 0$ corresponding to $\tilde{m}\to \infty$ and $\tilde{\Lambda}\to \infty$ with a fixed $\tilde{m}/\tilde{\Lambda}$, the chiral condensate is dominated by the UV divergent part,
\begin{align}
    \langle{\mathrm{BD}}|\bar{\psi}\psi|{\mathrm{BD}}\rangle \simeq \langle \bar{\psi}\psi\rangle_\Lambda = -\dfrac{m}{4\pi^2}\left(\Lambda^2e^{-m^2/\Lambda^2}-m^2\Gamma(0,m^2/\Lambda^2)\right)+\mathcal{O}(H^2).\label{eq:chiral_condensate_flat_space}
\end{align}
In the opposite limit $H\to \infty$ corresponding to $\tilde{m}\to 0$ and $\tilde{\Lambda}\to 0$ with a fixed $\tilde{m}/\tilde{\Lambda}$, both UV convergent and divergent parts contribute to the chiral condensate at the same order and we find
\begin{align}
    &\langle{\mathrm{BD}}|\bar{\psi}\psi|{\mathrm{BD}}\rangle  = -\frac{H^3\tilde{m}}{4\pi^2}\left(1-2\gamma_E-2\log\tilde{m}-\Gamma(0,\tilde{m}^2/\tilde{\Lambda}^2)\right) + \mathcal{O}(H^3\tilde{m}^3\log\tilde{m}),\label{eq:chiral_condensate_large_H}
\end{align}
where $\gamma_E$ denotes the Euler--Mascheroni constant.

We next consider an extremely large magnetic field, $e\tilde{B}\gg \tilde{\Lambda}^2$ and $e\tilde{B}\gg 1$.
In this limit all nonzero modes with $n\geq 1$ become negligible in the convergent part.
According to the detailed calculations provided in Appendix~\ref{app:chiral_condensate}, the regularized and convergent parts of the condensate in this large-magnetic-field limit can be analytically evaluated as
\begin{align}
\langle\bar{\psi}\psi\rangle_{\Lambda}
&=
-\frac{m eB_{\mathrm{phys}}}{4\pi^{2}}
\Gamma\left(0,\frac{m^{2}}{\Lambda^{2}}\right)+\mathcal{O}(e^{-2e\tilde{B}/\tilde{\Lambda}^2}),\\*
\langle\bar{\psi}\psi\rangle_{\mathrm{con}}
&=
\frac{H^{3}e\tilde{B}}{4\pi^{2}}
\left[
-2\tilde{m}\log\tilde{m}
+\tilde{m}
\left\{
\psi^{(0)}(-i\tilde{m})+\psi^{(0)}(i\tilde{m})
\right\}
\right]
+\mathcal{O}\left(\frac{1}{e\tilde{B}}\right).
\end{align}

When expressed in physical units, the cutoff-dependent part of
$\langle \bar{\psi}\psi\rangle_{\Lambda}$ coincides with that in the flat-space limit.
In the large-magnetic-field limit, its cutoff dependence is logarithmic,
\begin{align}
\Gamma(0,\tilde{m}^2/\tilde{\Lambda}^2)
=
-\gamma_E
-2\log(\tilde{m}/\tilde{\Lambda})
+\mathcal{O}(\tilde{m}^2/\tilde{\Lambda}^2),
\end{align}
in contrast to the weak magnetic field regime, where the cutoff dependence is quadratic.
This logarithmic behavior reflects the dominance of the lowest Landau level.
For $eB_{\mathrm{phys}}\gg m^2$, higher Landau levels are separated by an energy gap of order
$\sqrt{eB_{\mathrm{phys}}}$, and their contributions to the chiral condensate are suppressed.
As a result, the dynamics of charged fermions is effectively restricted to the direction parallel to the magnetic field.
In flat space at zero temperature, the lowest Landau level enhances the
infrared instability in the presence of an attractive interaction, thereby
amplifying the chiral condensate
\cite{Gusynin:1994xp,Gusynin:1994re,Gusynin:1994va,Gusynin:1995nb}.
This universal enhancement of chiral symmetry breaking by a magnetic field is
known as magnetic catalysis.
We will see that this mechanism also operates in de Sitter space when the Hubble parameter is sufficiently small.

\KFmod{Here we clarify the status of these limiting regimes within the cutoff theory. The NJL model can be regarded as a controlled local
effective theory only when the characteristic curvature and magnetic
scales satisfy $H\ll\Lambda$ and
$\sqrt{eB_{\mathrm{phys}}}\ll\Lambda$, respectively. The limits
$H/\Lambda\to\infty$ and $eB_{\mathrm{phys}}/\Lambda^2\to\infty$ considered
above and in Sec.~\ref{sec:analytic_phase_structure} should therefore be understood as formal diagnostic limits of the
regulated free-fermion bilinear. They are useful for isolating,
respectively, the large-curvature behavior and the lowest-Landau-level
dimensional reduction, but not for obtaining quantitatively controlled
predictions of the contact-interaction model. A quantitative treatment
in these regimes would require the higher-dimensional operators or the
UV completion of the NJL interaction to be specified.
Moreover, neglecting the effects of the magnetic energy density and anisotropic stress on the background requires $B_{\mathrm{phys}}^2\ll M_{\mathrm{pl}}^2H^2$, where $M_{\mathrm{pl}}$ is the reduced Planck mass.}

\subsection{Nambu--Jona-Lasinio model}\label{sec:njl_model}

Finally, we introduce the Nambu--Jona-Lasinio (NJL) model for a single Dirac fermion and derive the gap equation.
Solving the gap equation (in the mean-field approximation) determines whether a fermion mass is dynamically generated.

In the NJL model, the path integral is defined by
\begin{align}
Z
=
\int D\bar{\psi}D\psi\,e^{iS[\bar{\psi},\psi]},~
S[\bar{\psi},\psi]
=
\int \mathrm{d}^4x \sqrt{-g}
\left\{
i\bar{\psi}\hat{\gamma}^\mu D_\mu\psi
+
\dfrac{G}{2}
\left[
(\bar{\psi}\psi)^2
+
(\bar{\psi}i\gamma_5\psi)^2
\right]
\right\}.
\label{eq:njl_action}
\end{align}
Here, $G>0$ is the four-Fermi coupling.
The boundary condition of the path integral is chosen to be the Bunch--Davies vacuum.
Strictly speaking, the path integral in de Sitter space should be formulated in the in-in formalism~\cite{Weinberg:2005vy}.
However, the above notation is sufficient for the present purpose since we only use the two-point function in the free theory.
The classical action is invariant under the $U(1)_V\times U(1)_A$ transformation parametrized by real parameters $\alpha$ and $\beta$:
\begin{align}
U(1)_V:\quad
\psi \to e^{i\alpha}\psi,
\qquad
U(1)_A:\quad
\psi \to e^{i\beta\gamma_5/2}\psi .
\end{align}

Our computation follows ref.~\cite{Tong:2023krn}, except that we include an external magnetic field and use a different regularization scheme.
Introducing auxiliary fields $\sigma$ and $\pi_a$, the path integral can be rewritten exactly as
\begin{align}
Z
=
\int D\bar{\psi}D\psi D\sigma D\pi_a\,
e^{iS_{\mathrm{NJL}}},
\end{align}
where
\begin{align}
S_{\mathrm{NJL}}
=
\int \mathrm{d}^4x \sqrt{-g}
\left[
\bar{\psi}
\left(
i\hat{\gamma}^\mu D_\mu
-\sigma
-i\gamma_5\pi_a
\right)
\psi
-
\dfrac{\sigma^2+\pi_a^2}{2G}
\right].
\end{align}
Integrating out $\sigma$ and $\pi_a$ reproduces Eq.~\eqref{eq:njl_action}.

In our analysis, we adopt the mean-field approximation,
\begin{align}
\sigma(x)=\sigma,
\qquad
\pi_a(x)=\pi_a .
\end{align}
We parametrize the auxiliary fields as
\begin{align}
\sigma=\Delta\cos\theta,
\qquad
\pi_a=\Delta\sin\theta,
\qquad
\Delta^2=\sigma^2+\pi_a^2 .
\end{align}
Then the fermion mass term is written as
\begin{align}
\sigma+i\gamma_5\pi_a
=
\Delta e^{i\theta\gamma_5}.
\end{align}
The phase $\theta$ can be removed by a constant chiral rotation. After
this transformation, the fermionic action becomes
\begin{align}
S_{\Delta}
=
\int \mathrm{d}^4x\sqrt{-g}
\left[
i\bar{\psi}\hat{\gamma}^{\mu}D_{\mu}\psi
-
\Delta\bar{\psi}\psi
\right].
\end{align}
This is precisely the action of a free massive Dirac fermion with mass
$m=\Delta$ in the same de Sitter and magnetic-field background.
Therefore, within the mean-field approximation, the free-theory
fermion bilinear derived in Secs.~\ref{sec:charged_fermion_uniform_B} and~\ref{sec:chiral_condensate_limits} can be directly applied
by making the replacement $m\to\Delta$. In this chiral frame, the effective potential depends only on $\Delta$:
\begin{align}
V_{\mathrm{eff}}(\Delta)
=
\dfrac{i\log Z_{\Delta}}{\int \mathrm{d}^4x\sqrt{-g}}
+
\dfrac{\Delta^2}{2G},
\end{align}
where
\begin{align}
Z_{\Delta}
=
\int D\bar{\psi}D\psi\,e^{iS_{\Delta}} .
\end{align}

The gap equation is therefore
\begin{align}
\dfrac{\Delta_*}{G}
=
-
\left.
\langle {\mathrm{BD}}|\bar{\psi}\psi|{\mathrm{BD}}\rangle
\right|_{\Delta=\Delta_*}.
\label{eq:gap_equation}
\end{align}
A nonzero solution $\Delta_*\neq0$ gives the dynamically generated fermion mass.
\KFmod{Here the bilinear is evaluated at a fixed time using the mode functions for a constant mass $m=\Delta_*$.}
In practice, we evaluate
$\langle {\mathrm{BD}}|\bar{\psi}\psi|{\mathrm{BD}}\rangle$
using Eq.~\eqref{eq:renormalized_chiral_condensate}.
For generic parameter values, we solve the gap equation numerically and present the results in the next section.

\section{Phase diagram in de Sitter space with magnetic field}\label{sec:phase_diagram}

In this section, we discuss the phase diagram of the chiral condensate in an external magnetic field in de Sitter space.
We first present the phase diagram within the framework of the NJL model, and then discuss the competing effects of the magnetic field and spacetime curvature.
The phase diagrams below show fixed-time mean-field results obtained by evaluating the gap equation at a given $\eta$ with the corresponding $B_{\mathrm{phys}}(\eta)$ treated as a fixed parameter.
In this section and in the figures, $B$ denotes the physical magnetic field $B_{\mathrm{phys}}$ when it appears in dimensionless combinations such as $eB/\Lambda^2$.

\subsection{Analytic result: qualitative behavior of the phase structure}\label{sec:analytic_phase_structure}

In this subsection, we investigate the overall behavior of the phase structure in the theory defined by \eqref{eq:njl_action}.
Because the small- and large-magnetic-field limits significantly simplify the expression for $\langle{\mathrm{BD}} |\bar{\psi}\psi |{\mathrm{BD}}\rangle$, one can discuss the phase structure in these limits analytically.

Within our setup, the onset of chiral condensation can be captured by the following condition:
\begin{align}
    \left.\dfrac{\mathrm{d}^2 V_{\mathrm{eff}}}{\mathrm{d}\Delta^2}\right|_{\Delta=0}
    =
    \dfrac{1}{G}
    +
    \left.
    \dfrac{\mathrm{d}\langle {\mathrm{BD}}|{\bar{\psi}\psi}|{\mathrm{BD}}\rangle}{\mathrm{d}\Delta}
    \right|_{\Delta=0}
    <0 .
\label{eq:general_condensation_condition}
\end{align}
It is apparent that this condition identifies the parameter region in which the solution of the gap equation at $\Delta=0$ is locally unstable.
Strictly speaking, this condition is not sufficient to identify the true condensed phase if a first-order phase transition occurs, since several local minima may coexist. Nevertheless, as will be confirmed in the numerical analysis below, no such first-order transition appears in our case.
Therefore, the above condition provides a useful way to estimate the parameter region in which condensate formation can occur.

\subsubsection{The vanishing magnetic field limit}

\begin{figure}[t]
\centering
\includegraphics[width=0.4\textwidth]{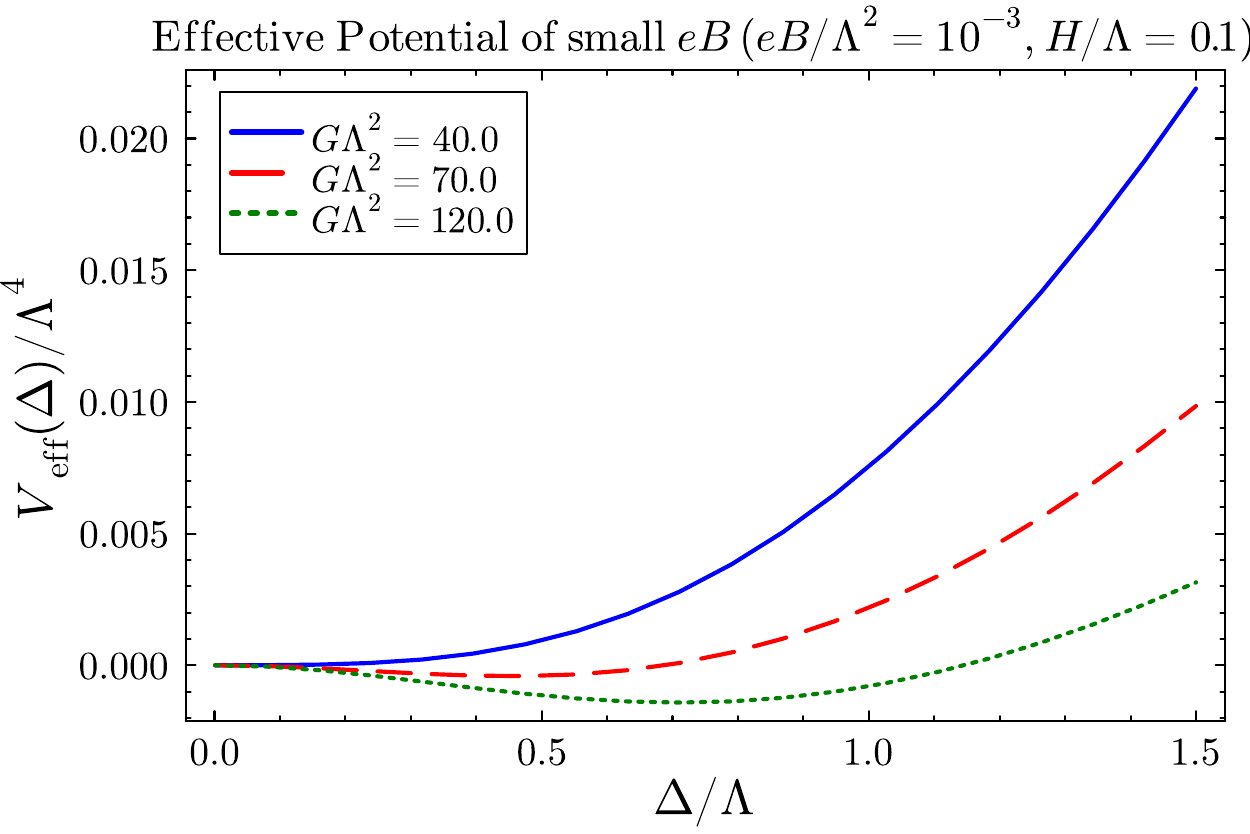}
\qquad
\includegraphics[width=0.4\textwidth]{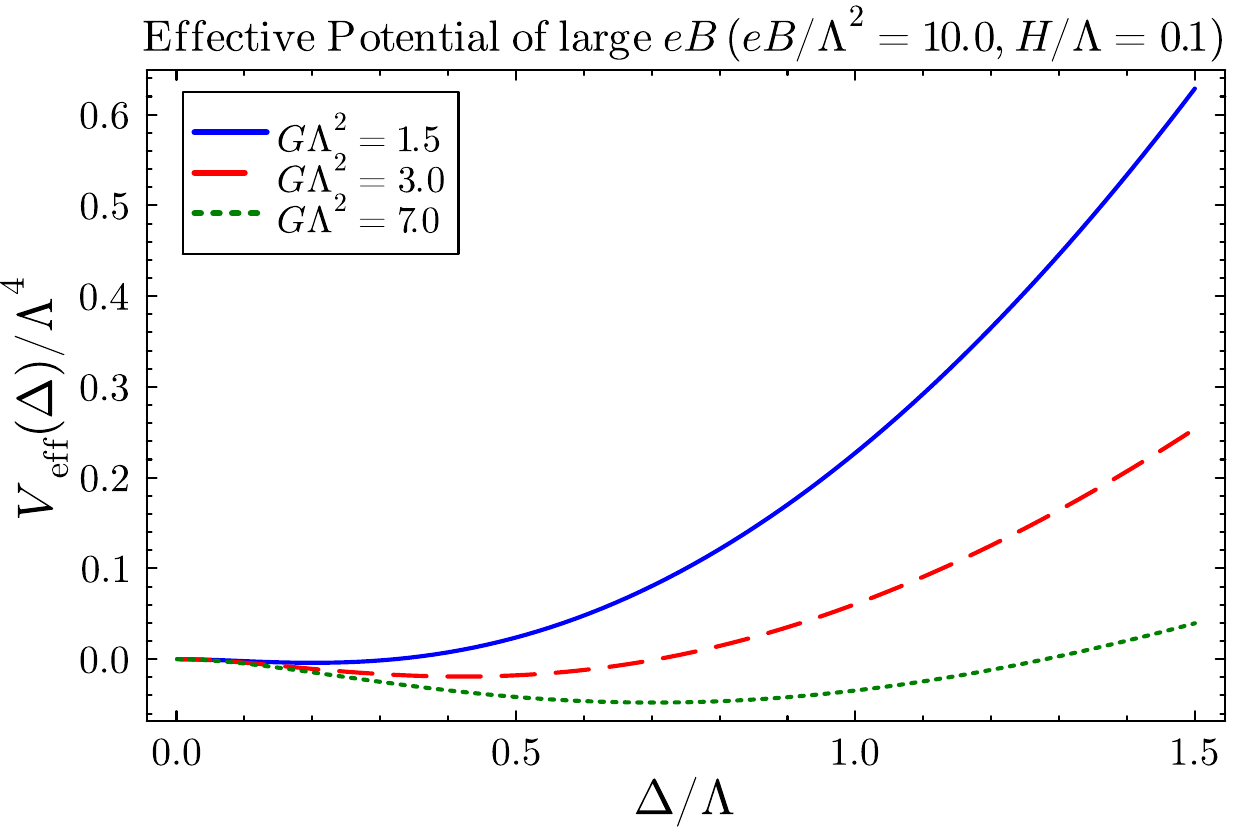}
\caption{Effective potentials normalized by the cutoff scale $\Lambda$ for small $eB/\Lambda^2=10^{-3}$ (left panel) and large $eB/\Lambda^2=10$ (right panel) magnetic fields are shown with $H/\Lambda =0.1$ for several values of the four-Fermi coupling.
}\label{fig:analytic_potentials}
\end{figure}

We consider the phase behavior in the limit of a vanishing magnetic field, $e\tilde{B}\to 0$. 
From the results in Sec.~\ref{sec:chiral_condensate_limits}, the convergent and regularized contributions to the chiral condensate are given by Eqs.~\eqref{eq:vanishing_magnetic_field_limit_convergent} and \eqref{eq:vanishing_magnetic_field_limit_divergent}, respectively, where the Dirac mass $\tilde{m}=m/H$ is replaced by the gap $\tilde{\Delta}=\Delta/H$.
Figure~\ref{fig:analytic_potentials} illustrates the qualitative behavior of the effective potential in the limiting regimes discussed below.

Expanding these expressions up to the linear order in $\Delta$, and using the asymptotic behaviors $\Gamma(0,x)=-\gamma_E-\log x+\mathcal{O}(x),~\psi^{(0)}(-1+x)=-1/x+1-\gamma_E+\mathcal{O}(x)$, one obtains the following expressions:
\begin{align}
\begin{split}
\langle\bar{\psi}\psi \rangle_{\mathrm{con}}&= \frac{H^2 \Delta}{4\pi^2} \left(2\log\frac{\Delta}{H} + 2\gamma_E - 1 \right)+\mathcal{O}(\Delta^2),\\
\langle \bar{\psi}\psi\rangle_\Lambda &=  -\frac{\Delta H^2}{4\pi^2}\left( \dfrac{\Lambda^2}{H^2} + \gamma_E + 2\log\frac{\Delta}{\Lambda} \right) + \mathcal{O}(\Delta^2).
\end{split}
\end{align}
Therefore, the total chiral condensate is evaluated as
\begin{align}
\begin{split}
\langle\bar\psi\psi\rangle_{\mathrm{total}} 
&= \langle\bar\psi\psi\rangle_\Lambda + \langle\bar\psi\psi\rangle_{\text{con}} \\
&= \frac{\Delta H^2}{4\pi^2} \left( -\tilde{\Lambda}^2 +  2\log\tilde{\Lambda} + \gamma_E - 1  \right)+ \mathcal{O}(eB_{\mathrm{phys}})+\mathcal{O}(\Delta^2).
\end{split}
\end{align}
Using this expression, the condition for the onset of condensate formation, Eq.~\eqref{eq:general_condensation_condition}, can be written explicitly as
\begin{equation}
\frac{GH^2}{4\pi^2} \left( \tilde{\Lambda}^2 - 2\log\tilde{\Lambda} - \gamma_E +1 \right) > 1. \label{eq:condensation_condition_vanishing_B}
\end{equation}
In the limit $\tilde{\Lambda}\to \infty$, the standard flat-spacetime result, $G\Lambda^2 > 4\pi^2$, is recovered.
Also notice that the expression inside the parentheses has a minimum at $\tilde{\Lambda}=1$. This nonmonotonic behavior may be interpreted as an artifact of our regularization scheme. Assuming that the cutoff scale is higher than the Hubble scale, the following analysis focuses on the regime $\tilde{\Lambda}>1$, in which a larger value of $GH^2$ is required as we lower $\tilde{\Lambda}$. Physically, this means that a stronger attractive force is required for condensation as we increase the Hubble temperature.

We next consider the solution of the gap equation for a nonzero $\Delta$ under the condition \eqref{eq:condensation_condition_vanishing_B}.
The gap equation is given by
 \begin{align}
\frac{4\pi^2}{GH^2}
&=
\tilde{\Lambda}^2
e^{-\tilde{\Delta}_*^2/\tilde{\Lambda}^2}
-
(1+\tilde{\Delta}_*^2)
\Gamma\left(0,\frac{\tilde{\Delta}_*^2}{\tilde{\Lambda}^2}\right)
\nonumber\\
&\quad
-
\left[
1
+2(1+\tilde{\Delta}_*^2)\log\tilde{\Delta}_*
-
(1+\tilde{\Delta}_*^2)
\left\{
\psi^{(0)}(-1-i\tilde{\Delta}_*)
+
\psi^{(0)}(-1+i\tilde{\Delta}_*)
\right\}
\right].
\end{align}
The solution of the gap equation can be evaluated analytically in the limit $\tilde{\Delta}_*\ll 1$:
\begin{align}
\Delta_*
&\simeq
H
\left[
\frac{
\tilde{\Lambda}^2
-2\log\tilde{\Lambda}
-\gamma_E
+1
-\dfrac{4\pi^2}{GH^2}
}{
1+\gamma_E
+2\log\tilde{\Lambda}
+\dfrac{1}{\tilde{\Lambda}^2}
-2\zeta(3)
}
\right]^{1/2}.
\end{align}
The square-root behavior shows that the gap $\Delta_*$ is nonanalytic at the critical point $\Delta_*=0$.
The phase transition is therefore second order when the Hubble parameter $H$ is varied with fixed $G$ and $\Lambda$.

One can also consider the flat-space limit $eB_{\mathrm{phys}}\to0$, $H\to0$, $0<1-4\pi^2/(G\Lambda^2)\ll1$, which gives the following expression:
\begin{align}
\Delta_*
&\simeq
\Lambda
\left[
-
\frac{
1-\dfrac{4\pi^2}{G\Lambda^2}
}{
W_{-1}
\left(
-e^{\gamma_E-1}
\left[
1-\dfrac{4\pi^2}{G\Lambda^2}
\right]
\right)
}
\right]^{1/2}.
\end{align}
Here, $W_{-1}(z)$ is the Lambert $W$ function defined by
\begin{align}
W_{-1}(z)e^{W_{-1}(z)}
&=
z,
\qquad
-\,e^{-1}\leq z<0,
\qquad
W_{-1}(z)\leq -1.
\end{align}
This is the standard result in the NJL model in flat space without a magnetic field.

\subsubsection{The large-magnetic-field limit}

We consider the phase behavior in the large-magnetic-field limit $eB_{\mathrm{phys}}\to\infty$.
The divergent and convergent parts of the chiral condensate are given by
\begin{align}
\langle\bar{\psi}\psi\rangle_\Lambda 
&= -\frac{H^3e\tilde{B}\tilde{\Delta}}{4\pi^2} \Gamma\left(0,\tilde{\Delta}^2/\tilde{\Lambda}^2\right) +\mathcal{O}(e^{-2e\tilde{B}/\tilde{\Lambda}^2}),\\
\langle\bar{\psi}\psi\rangle_{\mathrm{con}} 
&= \frac{H^3e\tilde{B}}{4\pi^2}\left(- 2\tilde{\Delta}\log \tilde{\Delta} + \tilde{\Delta}\bigl(\psi^{(0)}(-i\tilde{\Delta})+\psi^{(0)}(i\tilde{\Delta})\bigr) \right).
\end{align}
Using the asymptotic behaviors $\Gamma(0,x)=-\gamma_E-\log x+ \mathcal{O}(x)$ and $\psi^{(0)}(\pm ix)=\pm i/x-\gamma_E+\mathcal{O}(x)$ yields the following result:
\begin{align}
\begin{split}
\langle\bar\psi\psi\rangle_{\mathrm{total}} 
= -\dfrac{H^3e\tilde{B}}{4\pi^2}\left( 2\log \tilde{\Lambda} + \gamma_E \right)\tilde{\Delta} +\mathcal{O}(\tilde{\Delta}^2)+\mathcal{O}\left(\frac{1}{e\tilde{B}}\right).
\end{split}
\end{align}
Therefore, the condition in Eq.~\eqref{eq:general_condensation_condition} turns out to be
\begin{align}
\dfrac{GH^2e\tilde{B}}{4\pi^2}\left( 2\log \tilde{\Lambda} + \gamma_E \right)>1. \label{eq:condensation_condition_large_B}
\end{align}
It is useful to compare this with the numerical solution of the gap equation without the large-magnetic-field approximation.
We will see in Fig. \ref{fig:phase_boundary_large_magnetic_field} that the numerical and analytical results agree in the large-magnetic-field limit.

An important observation here is that the condition \eqref{eq:condensation_condition_large_B} can be satisfied even if $G\Lambda^2\to 0$, as long as $GeB_{\mathrm{phys}}\gg 1$ with $\tilde{\Lambda}\geq 1$.
Hence a nonzero condensate becomes a solution of the gap equation even when the attraction is very small compared with the cutoff scale, unlike the case without a magnetic field.

To quantify the size of the gap in the condensed phase, we can solve the following gap equation in the large-magnetic-field limit:

\begin{align}
\label{eq:gap_equation_large_magnetic_field}
\frac{4\pi^2}{G eB_{\mathrm{phys}}}
&=
\Gamma\left(0,\frac{\Delta_*^2}{\Lambda^2}\right)
+
2\log\frac{\Delta_*}{H}
-
\psi^{(0)}\left(i\frac{\Delta_*}{H}\right)
-
\psi^{(0)}\left(-i\frac{\Delta_*}{H}\right).
\end{align}
A solution in the small gap limit $\Delta_*/H\ll 1$ with finite $H$ is given by
\begin{align}
\Delta_*
&\simeq
H
\left[
\frac{
2\log\dfrac{\Lambda}{H}
+
\gamma_E
-
\dfrac{4\pi^2}{G eB_{\mathrm{phys}}}
}{
2\zeta(3)
-
\dfrac{H^2}{\Lambda^2}
}
\right]^{1/2},\\
0
&<
2\log\frac{\Lambda}{H}
+
\gamma_E
-
\frac{4\pi^2}{G eB_{\mathrm{phys}}}\ll
2\zeta(3)
-
\frac{H^2}{\Lambda^2}.
\end{align}
The above result shows that a second-order phase transition occurs when the magnetic field strength $B_{\mathrm{phys}}$ is varied with fixed $H$ and $G$.
Taking $H\to 0$ directly in Eq.~\eqref{eq:gap_equation_large_magnetic_field}, one obtains
\begin{align}\label{eq:gap_flat_limit}
\Delta_*
=
e^{-\gamma_E/2}
\Lambda
\exp\left[
-\frac{2\pi^2}{G eB_{\mathrm{phys}}}
\right],~\dfrac{\Delta_*}{\Lambda} \ll 1.
\end{align}
Although the gap is induced by the large magnetic field even in the weak-coupling limit $G\to 0$, its size is exponentially suppressed.
This exponentially small gap induced by the magnetic field for a weak attractive interaction $G\Lambda^2\ll 1$ is consistent with the original analysis~\cite{Gusynin:1994xp}.

\subsection{Numerical results: phase diagram}
\label{sec:numerical_phase_diagram}

\begin{figure}[t]
\centering
\includegraphics[width=.4\textwidth]{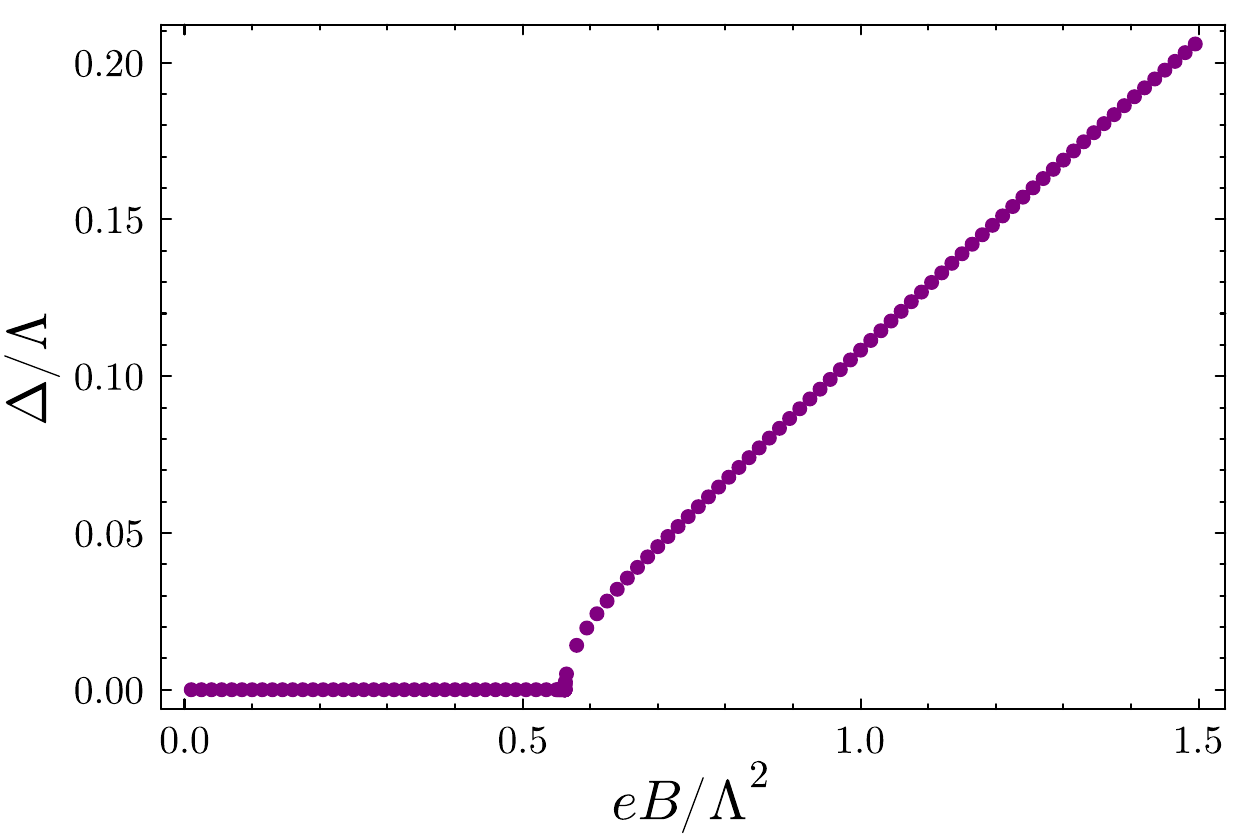}
\qquad
\includegraphics[width=.4\textwidth]{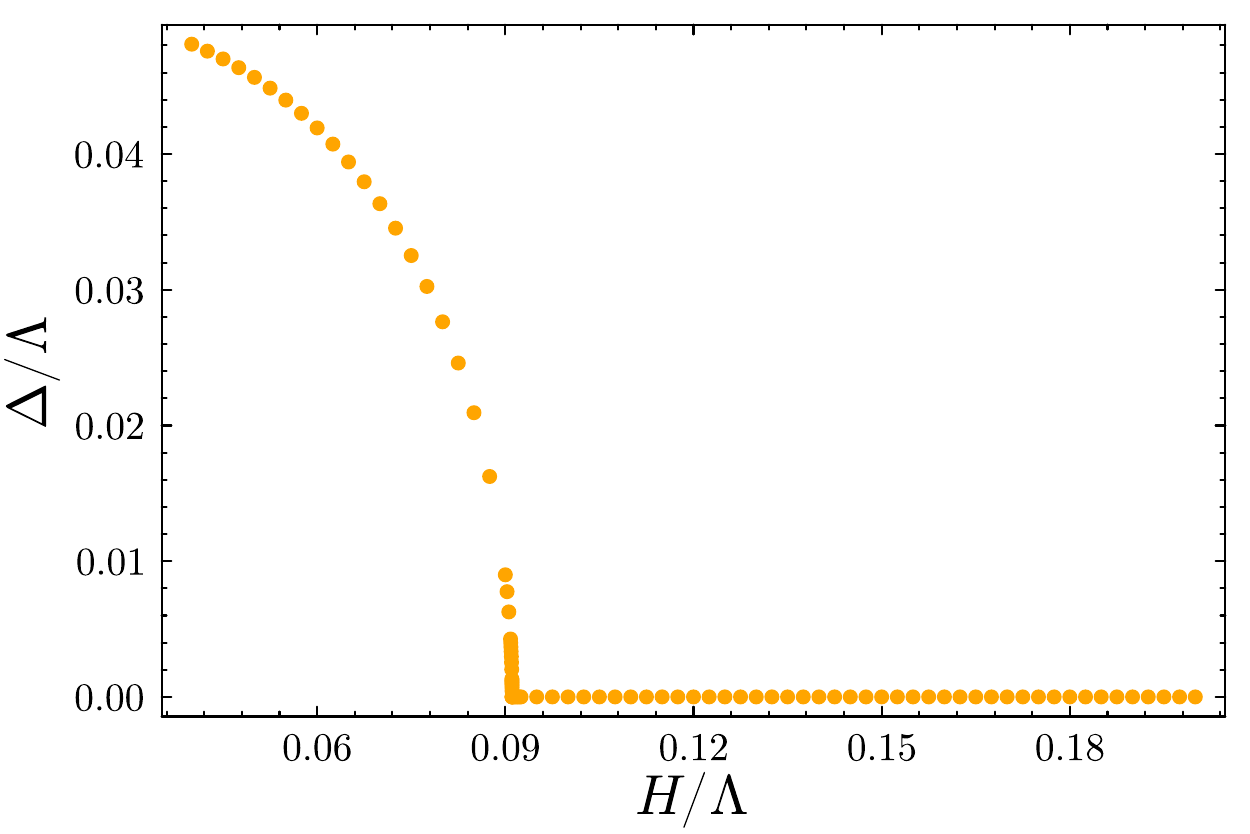}
\caption{The dependence of the chiral condensate on $eB/\Lambda^2$ at $H/\Lambda=0.05$ (left) and on $H/\Lambda$ at $eB/\Lambda^2=0.7$ (right), both for $G\Lambda^2=10$.
}\label{fig:gap_dependence}
\end{figure}

In this subsection, we present the phase diagram of the chiral condensate in an external magnetic field in de Sitter space, obtained numerically.

Let us first describe the numerical algorithm used to determine the phase structure.
In the numerical calculation, physical quantities are normalized by the cutoff scale $\Lambda$ as
\begin{align}
\hat{\Delta}&=\frac{\Delta}{\Lambda},
    \quad
    \hat{H}=\frac{H}{\Lambda},
    \quad
    e\hat{B}=\frac{eB}{\Lambda^2},
    \quad
    \hat{G}=G\Lambda^2.
\end{align}

Therefore, our results are invariant under a common rescaling of dimensionful parameters, and the phase diagram can be applied to any mass scale.
With this normalization, the gap equation can be rewritten as
\begin{align}
    \hat{\Delta}_* + \hat{G} \frac{\langle \bar{\psi}\psi\rangle}{\Lambda^3}(\hat{\Delta}_*;\hat{H},e\hat{B}) = 0.\label{eq:dimensionless_gap_equation}
\end{align}
We solve the above equation for fixed values of $\hat{G}$, $\hat{H}$, and $e\hat{B}$ to find the gap $\hat{\Delta}_*$.

The numerical calculation of the finite part of the chiral condensate in Eq.~\eqref{eq:chiral_condensate_one_loop} involves an integral over the momentum $k_z$ and the sum over the Landau levels $n$.
We introduce truncations for both the integral and the sum to control the numerical accuracy.
A smaller $e\hat{B}$ requires a larger number of Landau levels, which is computationally expensive.
Furthermore, the region $\hat{H}<0.05$ is also computationally expensive because the Hankel functions become highly oscillatory in this region.
However, both regions are well understood from the analytic expressions discussed in the previous subsection.
For this reason, the numerical calculations are restricted to the appropriate parameter region.

To detect multiple solutions of Eq.~\eqref{eq:dimensionless_gap_equation}, we divide the interval $10^{-2}\leq\hat{\Delta}\leq 0.8$ into a fine grid and search for neighboring grid points across which the function changes sign.
Each sign-changing interval is then used as the initial interval for a bracketing root-finding solver, which determines the corresponding root with higher numerical accuracy.

Using this method, we find that no multiple solutions exist in the region we investigated.
Although no first-order phase transition is observed in our setup, continuous phase transitions occur where the solution of the gap equation continuously changes from $\hat{\Delta}_*=0$ to $\hat{\Delta}_*\neq 0$.
The phase boundary is identified as the set of points where the solution develops a nonzero value, $\Delta_*\neq 0$.
In practice, the bisection method is used to determine the phase boundary efficiently.

Figure~\ref{fig:gap_dependence} shows the dependence of the chiral condensate $\Delta$ on the magnetic field and the Hubble parameter.
In the left panel we vary the magnetic field strength at a fixed Hubble parameter, and in the right panel we vary the Hubble parameter at a fixed magnetic field strength. We perform a dense scan near the onset of the condensate to check continuity.
In both cases, the gap changes continuously, which implies that there is no first-order phase transition.


As the Hubble parameter increases, the chiral condensate decreases.
The chiral condensate vanishes at a certain critical value of the Hubble parameter, restoring chiral symmetry.
This behavior was also reported in previous studies without a magnetic field~\cite{Tong:2023krn,Fujikura:2025xgl}.
In comparison, as the magnetic field strength increases, the chiral condensate increases.
This behavior is also indicated by the left panel of the figure.
In flat space, this enhancement is known as magnetic catalysis~\cite{Gusynin:1994re,Gusynin:1994va}.
We confirm that magnetic catalysis also occurs in de Sitter space, at least in the present setup.
This result suggests that a large magnetic field generated during inflation could trigger chiral condensation.
However, there is a potential issue associated with the limited applicability of the NJL model, which will be discussed in Sec.~\ref{sec:conclusion}.
We confirm that the overall behavior of the chiral condensate mentioned above is unchanged for different values of the coupling constant $G$.

\begin{figure}[t]
\begin{center}
\includegraphics[width=100mm]{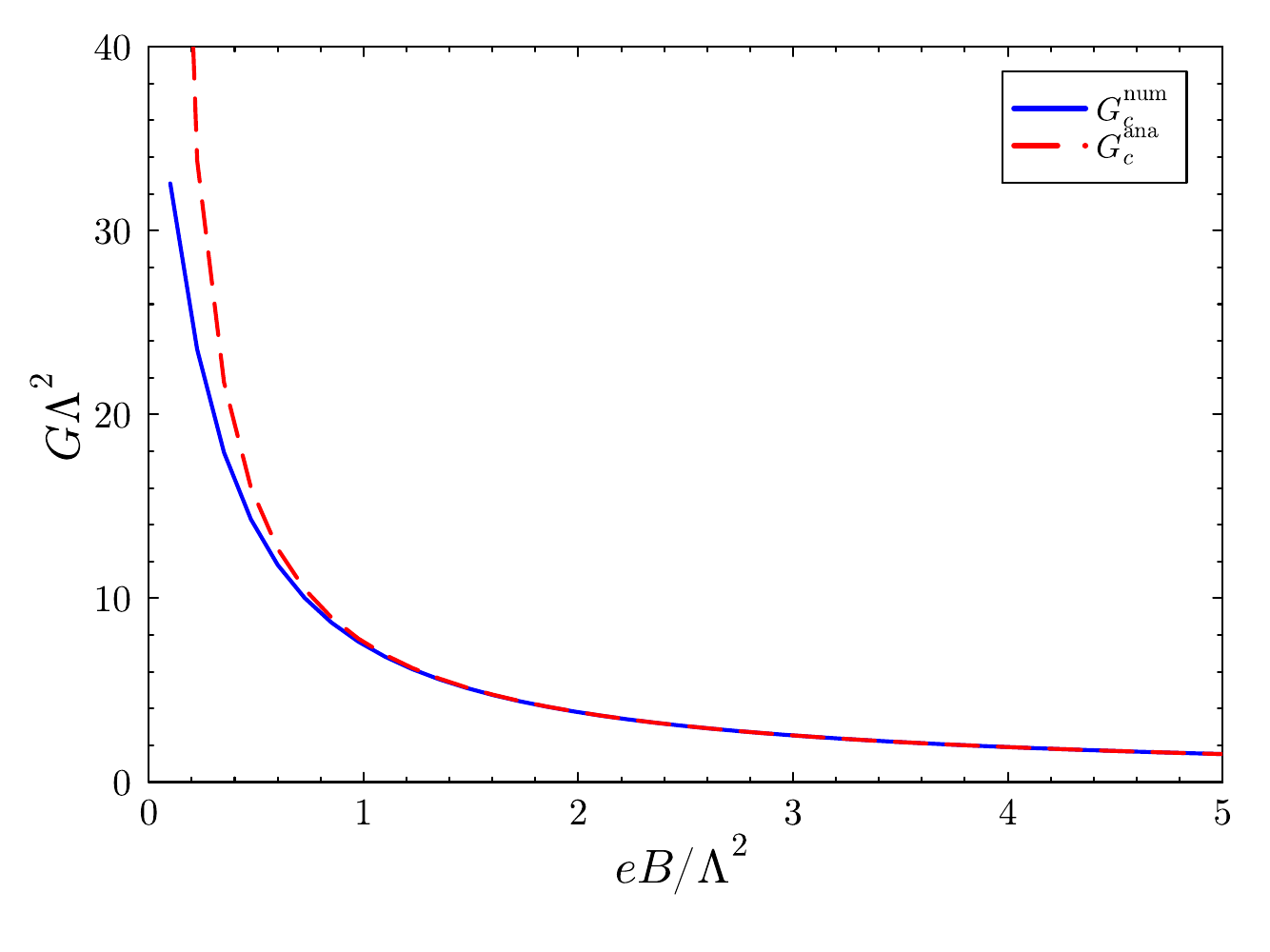}
\caption{The phase boundary is shown in the $(eB/\Lambda^2,G\Lambda^2)$ plane for $H/\Lambda =0.1$. The blue curve shows the numerical result, while the red dashed curve shows the analytic result relying on the large-magnetic-field limit, $eB_{\mathrm{phys}}\to\infty$.}\label{fig:phase_boundary_large_magnetic_field}
\end{center}
\end{figure}

Figure~\ref{fig:phase_boundary_large_magnetic_field} shows the phase boundary in the $(eB/\Lambda^2,G\Lambda^2)$ plane with a fixed Hubble parameter $H/\Lambda=0.1$.
The phase boundary is defined as the curve where the chiral condensate becomes nonzero. $G\Lambda^2$ represents the strength of the four-Fermi coupling. The blue solid line represents the numerically computed phase boundary, while the red dashed line corresponds to the analytical result shown in Eq.~\eqref{eq:condensation_condition_large_B}. The two agree well in the large-$B$ regime. Hence, this figure indicates that a larger coupling constant is required to trigger the chiral condensate when the magnetic field is weaker.
A remarkable point is that even when $G\Lambda^2\to 0$, the chiral condensate can be realized if the magnetic field is sufficiently strong.
This observation is important because the magnetic field generated during inflation can be parametrically larger than $H^2$, as discussed in Sec.~\ref{sec:conclusion}.

\begin{figure}[t]
\begin{center}
\includegraphics[width=100mm]{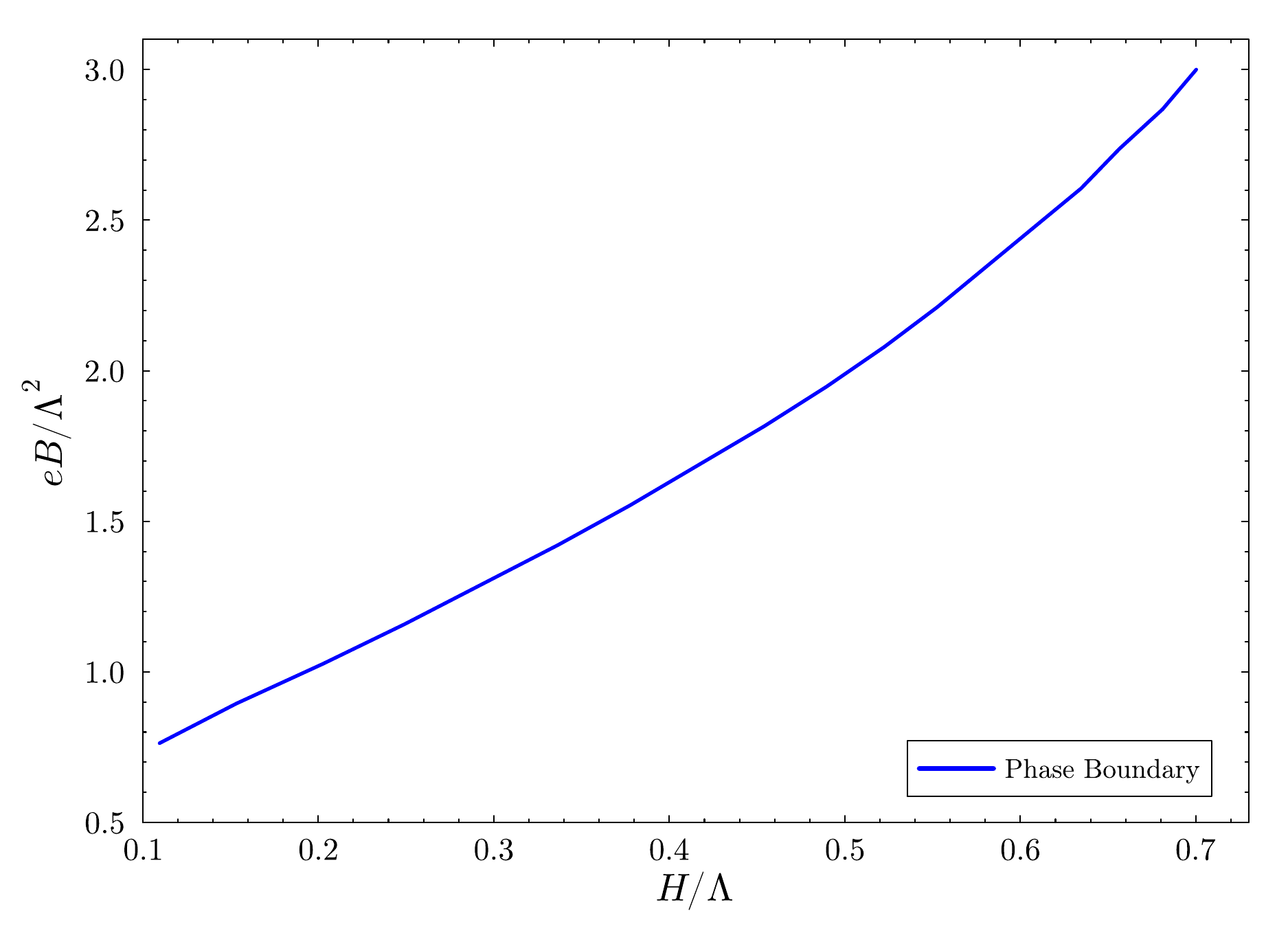}
\caption{The phase boundary with a magnetic field in the $(H/\Lambda,eB/\Lambda^2)$ plane with $G\Lambda^2=10$ is shown.
The blue solid line represents the phase boundary.}\label{fig:phase_boundary_H_B}
\end{center}
\end{figure}
Figure~\ref{fig:phase_boundary_H_B} shows the phase boundary between the condensed phase $\langle\bar{\psi}\psi\rangle\neq 0$ and the restored phase $\langle\bar{\psi}\psi\rangle=0$ in the $(H/\Lambda,eB/\Lambda^2)$ plane for $G\Lambda^2=10$.
For this parameter choice, there is no chiral condensate in the absence of the magnetic field, but the condensed phase is favored as the magnetic field increases, as can be seen from the figure.

\begin{figure}[t]
\begin{center}
\includegraphics[width=100mm]{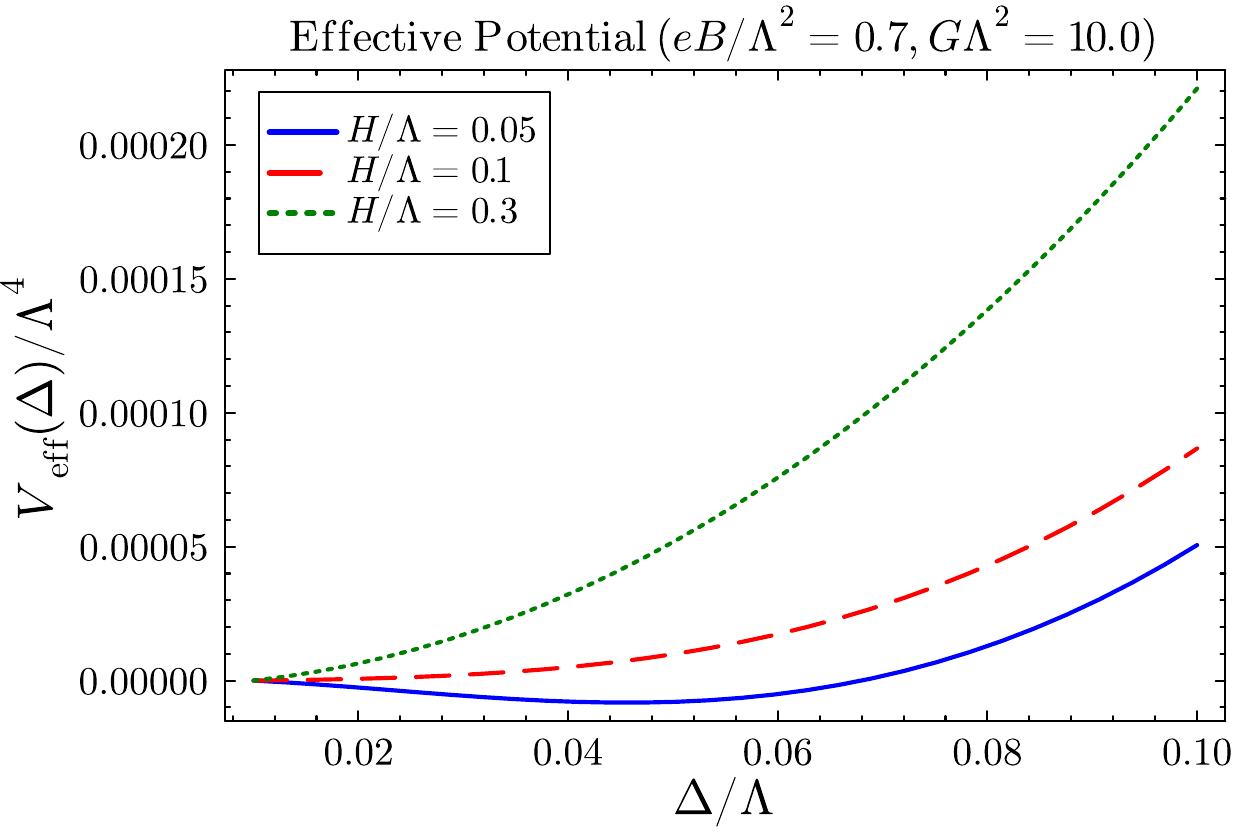}
\caption{The effective potentials normalized by the cutoff scale for different Hubble parameters, $H/\Lambda=0.05$ (blue curve), $H/\Lambda=0.1$ (red curve), and $H/\Lambda=0.3$ (green curve), with fixed $eB/\Lambda^2=0.7$ and $G\Lambda^2 = 10$ are shown.}\label{fig:potential}
\end{center}
\end{figure}
Figure~\ref{fig:potential} shows the effective potential for different values of the Hubble parameter.
We can see that the effective potential has a minimum at $\Delta\neq 0$ for a small value of the Hubble parameter.
As the Hubble parameter increases, the minimum shifts to the origin of the potential, and eventually $\Delta=0$ is realized, where chiral symmetry is restored.

\clearpage
\section{Conclusion and discussion}\label{sec:conclusion}

In this paper, we investigated the fate of chiral symmetry breaking in de Sitter space in the presence of a uniform background magnetic field. We discussed the phase structure using the NJL model. \SYmod{We found that an enhancement of the magnetic field is required for the chiral condensate to form in de Sitter space.}
An interesting observation is that a chiral condensate of charged fermions is induced by the magnetic field even when the four-Fermi coupling constant is very small. This effect is known as magnetic catalysis and has been extensively studied in flat space. In the present work, we confirmed that this phenomenon also occurs in de Sitter space when the magnetic field strength is much larger than $H^2$.
Spacetime curvature, on the other hand, suppresses the chiral condensate and eventually restores chiral symmetry in de Sitter space when the Hubble parameter exceeds a certain critical value.
Moreover, we obtained the phase diagram of this theory numerically.
\KFmod{In slow-roll inflation, we expect small corrections to the fixed-time phase diagram, with $H$ acting as a slowly varying parameter~\cite{Tong:2023krn,Fujikura:2025xgl}.}

A natural question is then whether chiral symmetry breaking occurs in the inflationary universe. In a recent study~\cite{Iarygina:2025schwinger}, magnetic field generation during the late stages of inflation was investigated using \KFmod{classical lattice computations}. For their benchmark parameter choice, the dimensionless physical magnetic field strength 
\SYmod{$\tilde{B}=B_{\mathrm{phys}}/H^2$ reaches as high as 
 $\tilde{B} \sim 10^1$ for $N_e=-2$ and 
 $\tilde{B} \sim 10^3$ for $N_e=0$, where $N_e$ is the $e$-folding number defined such that $N_e=0$ corresponds to the end of inflation.
Although the actual strength of the generated magnetic field strongly depends on the model parameters, the above results indicate that the magnetic field can be amplified during inflation and can reach values much larger than $H^2$ at the end of inflation.}
Furthermore, the correlation length of the magnetic field generated at the end of axion inflation is typically comparable to the Hubble scale $H^{-1}$~\cite{Adshead:2016iae}.
\KFmod{This motivates a locally homogeneous approximation to the magnetic field.}

\KFmod{In realistic inflationary settings, both the magnetic field and the condensate can evolve with time. A quantitative treatment requires solving the fermion mode equations in the evolving magnetic background and determining the condensate self-consistently from the resulting bilinear. The deviation from our fixed-time solutions remains to be quantified.}

The main limitation of this paper is the use of the NJL model.
In a realistic setup, the contact attractive interaction should be replaced by a nonlocal interaction, such as photon or gluon exchange.
In the weak-coupling regime in flat space, the Schwinger--Dyson equation in the ladder approximation reveals that the actual size of the gap is much smaller than that predicted by the NJL model~\cite{Gusynin:1995nb,Gusynin:1999pq,Miransky:2002rp}.
The NJL model assumes the strength of the attractive interaction to be constant, an assumption that is no longer justified when the magnetic field strength exceeds the cutoff scale.
The actual interaction strength is controlled by the running coupling constant at the magnetic scale, and the resulting gap is expected to be exponentially smaller than the NJL-model prediction.
Hence, our analysis provides only a qualitative picture of the phase structure, and a more sophisticated analysis is required to obtain quantitative results.
Therefore, the actual fermion gap induced by gauge interactions alone may be too small to be phenomenologically relevant.
If, however, an additional attractive interaction beyond the Standard Model gauge interactions is present, our analysis may provide a useful qualitative prediction once an appropriate matching condition is imposed.
It would be interesting to investigate soft excitations on top of the chiral condensate, since long-range correlations may be generated during inflation.

\acknowledgments
K.F. is supported by JST CREST Grant No. JPMJCR24I3 and JSPS KAKENHI Grant No. JP26K17144. T.N. is supported by JSPS KAKENHI Grant No. JP26K07101 and by the JGC-S Scholarship Foundation. S.Y. is supported by the World-Leading Innovative Graduate Study Program for Advanced Basic Science Course at the University of Tokyo. The authors used ChatGPT (OpenAI) and Claude (Anthropic) for language editing and for improving the readability of the manuscript. Claude and Codex (OpenAI) were also used to assist with proofreading and checking the manuscript, including checks of typographical errors and of the consistency of the notation, equations and bibliographic information. The authors reviewed and verified all AI-assisted output and take full responsibility for the content of this article.

\appendix
\section{Calculation of normalization constants}\label{app:normalization}
We calculate the normalization constant $C_n$. This constant is determined by requiring the field to satisfy the equal-time anticommutation relation:
\begin{equation}
    \lbrace\psi(\eta,\b{x}),\psi^\dagger(\eta,\b{x}')\rbrace = \frac{\delta^3(\b{x}-\b{x}')}{a^3(\eta)}.
\end{equation}
The field operator is expanded as
\begin{align}
    \psi(\eta,\b{x}) = \int\frac{\mathrm{d}^{2}\bar{\b{k}}}{(2\pi)^2} \sum_{n,s}\left[a_{\bar{\b{k}},s,n}U_{n,s,\bar{\b{k}}}(\eta,\b{x}) + b^{\dagger}_{\bar{\b{k}},s,n}V_{n,s,\bar{\b{k}}}(\eta,\b{x})\right].
\end{align}
Assuming the standard anticommutation relations for the creation and annihilation operators, the equal-time anticommutation relation is equivalent to the following orthonormality relations for the mode functions:
\begin{align}
\int \mathrm{d}^3\b{x}\,a^3(\eta) U_{n,s,\bar{\b{k}}}^\dagger(\eta,\b{x}) U_{n',s',\bar{\b{k}}'}(\eta,\b{x})
&=\int \mathrm{d}^3\b{x}\,a^3(\eta) V_{n,s,\bar{\b{k}}}^\dagger(\eta,\b{x}) V_{n',s',\bar{\b{k}}'}(\eta,\b{x}) \nonumber\\*
&= (2\pi)^2 \delta^2(\bar{\b{k}}-\bar{\b{k}}')\delta_{s,s'}\delta_{n,n'},\label{eq:orthonormality_1}\\*
\int \mathrm{d}^3\b{x}\,a^3(\eta) U_{n,s,\bar{\b{k}}}^\dagger(\eta,\b{x}) V_{n',s',\bar{\b{k}}'}(\eta,\b{x})
&=\int \mathrm{d}^3\b{x}\,a^3(\eta) V_{n,s,\bar{\b{k}}}^\dagger(\eta,\b{x}) U_{n',s',\bar{\b{k}}'}(\eta,\b{x})=0.\label{eq:orthonormality_2}
\end{align}

We now determine the normalization constant $C_n$ by substituting the mode functions of Eqs.~\eqref{eq:mode_function_1}--\eqref{eq:mode_function_4} into Eq.~\eqref{eq:orthonormality_1}. Note that Eq.~\eqref{eq:orthonormality_2} is automatically satisfied by these mode functions.
Since $C_n$ is a time-independent constant, it is sufficient to evaluate the integral in the early-time limit, $\eta \to -\infty$. 
In this limit, using the asymptotic expansion of the Hankel function, $\mathcal{H}_\nu^{(1)}(-\omega_n\eta)\sim\sqrt{\dfrac{2}{\pi}}e^{-i\omega_n\eta-i\pi\nu/2-i\pi/4}$, the spatial integration yields
\begin{equation}
\begin{split}
\int \mathrm{d}^3\b{x}\,a^3U_{n,s,\bar{\b{k}}}^\dagger U_{n',s',\bar{\b{k}}'}
&=\dfrac{2}{|C_n|^2 }\dfrac{2}{\pi}e^{-\pi m/H} (2\pi)^2 \delta^2(\bar{\b{k}}-\bar{\b{k}}')\delta_{s,s'}\delta_{n,n'}.
\end{split}
\end{equation}
During this calculation, the spatial integration over $x$ and $z$ produces the momentum delta function $(2\pi)^2 \delta^2(\bar{\b{k}}-\bar{\b{k}}')$, while the integration over $y$ uses the orthonormality of the Hermite functions:
\begin{equation}
\int^{\infty}_{-\infty}\mathrm{d}y\,h_N(\bar y)h_{N'}(\bar y)=\delta_{N,N'}.
\end{equation}
Comparing this result with the right-hand side of Eq.~\eqref{eq:orthonormality_1}, we obtain the condition
\begin{equation}
    \dfrac{1}{ |C_n|^2}\dfrac{4}{\pi}e^{-\pi m/H}=1.
\end{equation}
Choosing $C_n$ to be real and positive, we finally obtain
\begin{equation}
    C_n=\sqrt{\dfrac{4}{\pi}}e^{-\pi m/(2H)}.
\end{equation}

\section{Derivation of the chiral condensate in limiting regimes}
\label{app:chiral_condensate}

In this appendix, we derive analytic expressions for the chiral condensate in two limiting regimes: the vanishing- and large-magnetic-field limits.
We start with the expression for the chiral condensate at one-loop order, which is given by
\begin{equation}
\begin{aligned}
    &\langle {\mathrm{BD}}|\bar{\psi}(\eta,\b{x})\psi(\eta,\b{x}) |{\mathrm{BD}}\rangle_{\mathrm{con}}
    =
    H^3\int^\infty_{-\infty} \frac{\mathrm{d}\tilde{k}_z}{2\pi}
    \dfrac{e^{\pi \tilde{m}} e\tilde{B}}{4}
    \sum_{n=0}^{\infty}
    \mathcal{I}^{\mathrm{con}}_{n}(\tilde{k}_z,\tilde{B}),
    \\
    &\mathcal{I}^{\mathrm{con}}_n(\tilde{k}_z,\tilde{B})=
\begin{dcases}
\dfrac{\tilde{\omega}_{0}}{2}
\left(
|H^{(2)}_{\nu^*-1}(\tilde{\omega}_0)|^2
-
|H^{(2)}_{\nu^*}(\tilde{\omega}_0)|^2
\right)
-\mathcal{I}^{\mathrm{div}}_0
& (n=0),
\\[1.2ex]
\tilde{\omega}_n
\left(
|H^{(2)}_{\nu^*-1}(\tilde{\omega}_n)|^2
-
|H^{(2)}_{\nu^*}(\tilde{\omega}_n)|^2
\right)
-\mathcal{I}^{\mathrm{div}}_n
& (n\geq1),
\end{dcases}
\label{eq:chiral_condensate_one_loop_appendix}
\\
&\tilde{B}=B\eta^2,\qquad
\tilde{k}_z=-k_z\eta,\qquad
\tilde{\omega}_n^2=\tilde{k}_z^2 + 2ne\tilde{B},\qquad
\tilde{m}=m/H,
\\
&\nu=\frac{1}{2}-i\tilde{m}.
\end{aligned}
\end{equation}
Here, $\mathcal{I}^{\mathrm{div}}_n$ is the divergent part of the integrand, which is obtained from the large-argument asymptotic expansion of the Hankel function:
\begin{align}
\mathcal{I}^{\mathrm{div}}_n &=
\begin{dcases}
e^{-\pi \tilde{m}}
\left(
-\frac{2}{\pi}
\frac{\tilde{m}}{\sqrt{\tilde{k}_z^2+\tilde{m}^2}}
\right)
+\mathcal{O}\left(\frac{1}{\tilde{E}_n^5}\right)
& (n=0),
\\[1.2ex]
e^{-\pi \tilde{m}}
\left(
-\frac{4}{\pi}
\frac{\tilde{m}}{\sqrt{\tilde{\omega}_n^2+\tilde{m}^2}}
+
\frac{2}{\pi}
\frac{\tilde{m}}{(\tilde{\omega}_n^2+\tilde{m}^2)^{3/2}}
\right)
+\mathcal{O}\left(\frac{1}{\tilde{E}_n^5}\right)
& (n\geq 1),
\end{dcases}
\end{align}
where we defined $\tilde{E}_n^2
=
\tilde{k}_z^2+2ne\tilde{B}+\tilde{m}^2$ here.

\subsection{The vanishing magnetic field limit}

Let us first consider the vanishing magnetic field limit, $e\tilde{B}\to 0$.
In this limit, the sum over the Landau levels can be approximated by an integral over the transverse momentum $\tilde{k}_\perp$.
Defining $\tilde{\kappa}=2e\tilde{B} n$ and using the Euler--Maclaurin formula with the measure $\mathrm{d}\tilde{\kappa}=2e\tilde{B}$, the Landau-level sum in Eq.~\eqref{eq:chiral_condensate_one_loop_appendix} is approximated as
\begin{align}
    \sum^{\infty}_{n=0}
    e\tilde{B}\mathcal{I}_n(\tilde{k}_z,\tilde{B})
    \to
    \frac{1}{2}\int^\infty_0
    \mathrm{d}\tilde{\kappa}\,
    \tilde{\omega}
    \left(
    |H^{(2)}_{\nu^*-1}(\tilde{\omega})|^2
    -
    |H^{(2)}_{\nu^*}(\tilde{\omega})|^2
    \right),
    \qquad
    \tilde{\omega}^2=\tilde{k}_z^2+\tilde{\kappa}.
\end{align}
There is no $e\tilde{B}$ dependence in this limit.
$\tilde{\kappa}$ can be regarded as the square of the transverse momentum $\tilde{\kappa}=\tilde{k}_\perp^2$.
Hence, Eq.~\eqref{eq:chiral_condensate_one_loop_appendix} can be written as
\begin{align}
    \langle {\mathrm{BD}}|\bar{\psi}(\eta,\b{x})\psi(\eta,\b{x}) |{\mathrm{BD}}\rangle_{\mathrm{con}}
    &=
    \frac{H^3}{4\pi}
    \int^\infty_0\mathrm{d}\tilde{k}\,
    \Bigg[
    e^{\pi\tilde{m}}\tilde{k}^3
    \left(
    |H^{(2)}_{\nu^*-1}(\tilde{k})|^2
    -
    |H^{(2)}_{\nu^*}(\tilde{k})|^2
    \right)
    \nonumber\\
    &\hspace{20mm}
    -
    \tilde{k}^2
    \left(
    -\frac{4}{\pi}
    \dfrac{\tilde{m}}{\sqrt{\tilde{k}^2+\tilde{m}^2}}
    +
    \frac{2}{\pi}
    \dfrac{\tilde{m}}{(\tilde{k}^2+\tilde{m}^2)^{3/2}}
    \right)
    \Bigg].
\end{align}
The above expression is UV finite.

Consider the following deformed integrals:
\begin{equation}
\begin{aligned}
    F(q,\tilde{m})
    &\coloneqq
    e^{\pi\tilde{m}}
    \int^\infty_0\mathrm{d}z\,z^{q-1}
    \left(
    |H^{(2)}_{\nu^*-1}(z)|^2
    -
    |H^{(2)}_{\nu^*}(z)|^2
    \right),
    \\
    F_{\mathrm{asy}}(q,\tilde{m})
    &\coloneqq
    \int^\infty_0\mathrm{d}z\,z^{q-2}
    \left[
    -\frac{4}{\pi}
    \frac{\tilde{m}}{\sqrt{z^2+\tilde{m}^2}}
    +
    \frac{2}{\pi}
    \frac{\tilde{m}}{(z^2+\tilde{m}^2)^{3/2}}
    \right].
\end{aligned}
\end{equation}
The above integrals are convenient for extracting the UV divergence and the finite part of the chiral condensate.

We use the following relations:
\begin{equation}
\begin{aligned}
    &H^{(2)}_{\alpha}(z)
    =
    \dfrac{i}{\sin(\pi\alpha)}
    \left(
    -e^{i\pi\alpha}J_\alpha(z)+J_{-\alpha}(z)
    \right),
    \\*
    &H^{(1)}_{\beta}(z)
    =
    \dfrac{i}{\sin(\pi\beta)}
    \left(
    -J_{-\beta}(z)+e^{-i\pi\beta}J_\beta(z)
    \right),
    \\*
    &H^{(2)}_{\nu^*}(z)
    =
    \left(H^{(1)}_{\nu}(z)\right)^*\qquad(z\in\mathbb{R}_{>0}).
\end{aligned}
\end{equation}
Here, $J_\alpha(z)$ is the Bessel function of the first kind.

Using $\nu=\frac{1}{2}-i\tilde{m}$, one can decompose the integrand of $F(q,\tilde{m})$ into Bessel functions as
\begin{equation}
\begin{aligned}
    |H^{(2)}_{\nu^*-1}(z)|^2
    -
    |H^{(2)}_{\nu^*}(z)|^2
    &=
    H^{(2)}_{-\nu}(z)H^{(1)}_{-\nu^*}(z)
    -
    H^{(2)}_{\nu^*}(z)H^{(1)}_{\nu}(z)
    \\
    &=
    \frac{1}{\cosh^2(\pi\tilde{m})}
    \left[
    \left(1-e^{-2\pi\tilde{m}}\right)
    \left(
    J_\nu(z) J_{\nu^*}(z)
    -
    J_{-\nu}(z)J_{-\nu^*}(z)
    \right)
    \right.
    \\
    &\hspace{25mm}
    \left.
    +2ie^{-\pi\tilde{m}}
    \left(
    J_{-\nu}(z)J_{\nu^*}(z)
    -
    J_{\nu}(z)J_{-\nu^*}(z)
    \right)
    \right].
\end{aligned}
\end{equation}
Therefore we only need to consider the following integrals:
\begin{equation}
\label{eq:definition_Mab}
\begin{aligned}
    M_{ab}(q)
    &\coloneqq
    \int^\infty_0\mathrm{d}z\,z^{q-1}J_a(z)J_b(z)
    \\
    &=
    \dfrac{
    2^{q-1}\Gamma(1-q)\Gamma((a+b+q)/2)
    }{
    \Gamma((2+a-b-q)/2)
    \Gamma((2-a+b-q)/2)
    \Gamma((2+a+b-q)/2)
    },
    \\
    &q<1,\qquad
    \Re(a+b+q)>0.
\end{aligned}
\end{equation}
In this expression, $\Gamma$ is the gamma function.
This result can be analytically continued to the whole complex plane of $q$, \SYmod{except for simple poles that generate the UV divergence as $\epsilon = 4-q\to 0$.} $M_{ab}(q)$ admits a meromorphic continuation, whose poles are determined by the gamma functions in Eq.~\eqref{eq:definition_Mab}. In the present calculation, only the pole near $q=4$ is relevant.
Hence we arrive at the following expression,
\begin{equation}
\begin{aligned}
F(q,\tilde{m})
&\coloneqq
e^{\pi\tilde{m}}
\int^\infty_0\mathrm{d}z\,z^{q-1}
\left(
|H^{(2)}_{\nu^*-1}(z)|^2
-
|H^{(2)}_{\nu^*}(z)|^2
\right)
\\
&=
\frac{e^{\pi\tilde{m}}}{\cosh^2(\pi\tilde{m})}
\left[
\left(1-e^{-2\pi\tilde{m}}\right)
\left(
M_{\nu,\nu^*}(q)-M_{-\nu,-\nu^*}(q)
\right)
\right.
\\
&\hspace{20mm}
\left.
+
2ie^{-\pi\tilde{m}}
\left(
M_{-\nu,\nu^*}(q)-M_{\nu,-\nu^*}(q)
\right)
\right].
\end{aligned}
\end{equation}
\SYmod{Taking the limit $\epsilon\to 0$}, one extracts the UV divergence and the finite part of $F(q,\tilde{m})$ as
\begin{align}
    F(4-\epsilon,\tilde{m})
    =
    \frac{\tilde{m}}{\pi}(1+\tilde{m}^2)
    \left(
        \frac{2}{\epsilon}
        -1
        +\log 4
        -\psi^{(0)}(-1-i\tilde{m})
        -\psi^{(0)}(-1+i\tilde{m})
    \right)
    +\mathcal{O}(\epsilon).
\end{align}
Here, $\psi^{(0)}$ is the digamma function.
For the asymptotic part, the integral can be performed directly, giving
\begin{align}
    F_{\mathrm{asy}}(4-\epsilon,\tilde{m})
    &=
    \frac{2(\tilde{m}+\tilde{m}^3)}{\pi\epsilon}
    +
    \frac{\tilde{m}}{\pi}
    \Bigl[
    -2
    +
    \tilde{m}^2(-1+\log 4)
    +
    \log 4
    \nonumber\\
    &\hspace{35mm}
    -2(1+\tilde{m}^2)\log \tilde{m}
    \Bigr]
    +\mathcal{O}(\epsilon).
\end{align}
We finally obtain the following expression for the convergent contribution:
\begin{equation}
\begin{aligned}
\langle\bar{\psi}\psi\rangle_{\mathrm{con}}
&=
\frac{H^3}{4\pi}
\left[
F(4-\epsilon,\tilde{m})
-
F_{\mathrm{asy}}(4-\epsilon,\tilde{m})
\right]
\\
&=
\frac{H^3}{4\pi^2}
\tilde{m}
\Bigl[
1
+
2(1+\tilde{m}^2)\log\tilde{m}\\
&\hspace{25mm}
-
(1+\tilde{m}^2)
\bigl(
\psi^{(0)}(-1-i\tilde{m})
+
\psi^{(0)}(-1+i\tilde{m})
\bigr)
\Bigr],
\qquad
(e\tilde{B}\to 0).
\end{aligned}
\end{equation}

\subsection{The large-magnetic-field limit}

In the opposite limit $e\tilde{B}\to \infty$, the contribution from the higher Landau levels can be neglected, and only the lowest Landau level contributes to the chiral condensate.
It is explicitly given by
\begin{align}
\langle {\mathrm{BD}}|\bar{\psi}(\eta,\b{x})\psi(\eta,\b{x}) |{\mathrm{BD}}\rangle_{\mathrm{con}}
&=
H^3\int^\infty_{-\infty}
\frac{\mathrm{d}\tilde{k}_z}{2\pi}
\dfrac{e^{\pi \tilde{m}} e\tilde{B}}{8}
\Bigg[
|\tilde{k}_{z}|
\left(
|H^{(2)}_{\nu^*-1}(|\tilde{k}_z|)|^2
-
|H^{(2)}_{\nu^*}(|\tilde{k}_z|)|^2
\right)
\nonumber\\
&\hspace{35mm}
-
e^{-\pi \tilde{m}}
\left(
-\frac{4}{\pi}
\frac{\tilde{m}}{\sqrt{\tilde{k}_z^2+\tilde{m}^2}}
\right)
\Bigg].
\end{align}
Using the same technique as in the vanishing magnetic field case, we can perform the integral over $\tilde{k}_z$ and extract the finite part as
\begin{align}
    \langle {\mathrm{BD}}|\bar{\psi}(\eta,\b{x})\psi(\eta,\b{x})|{\mathrm{BD}}\rangle_{\mathrm{con}}
    =
    \frac{H^3 e\tilde{B}}{4\pi^2}\,
    \tilde{m}
    \left[
    \psi^{(0)}(i\tilde{m})
    +
    \psi^{(0)}(-i\tilde{m})
    -
    2\log\tilde{m}
    \right],
    \qquad
    (e\tilde{B}\to \infty).
\end{align}

These expressions are used to discuss the phase structure in the vanishing- and large-magnetic-field limits in the main text.

\bibliographystyle{JHEP}
\bibliography{ref}
\end{document}